\documentclass[pdflatex,sn-nature]{sn-jnl}

\usepackage{graphicx}%
\usepackage{multirow}%
\usepackage{amsmath,amssymb,amsfonts}%
\usepackage{amsthm}%
\usepackage{mathrsfs}%
\usepackage[title]{appendix}%
\usepackage{xcolor}%
\usepackage{textcomp}%
\usepackage{manyfoot}%
\usepackage{booktabs}%
\usepackage{listings}%

\usepackage{epstopdf}

\usepackage[normalem]{ulem}

\usepackage[utf8]{inputenc}
\usepackage[T1]{fontenc}
\hypersetup{
  pdftitle={Turning angle analysis reveals hidden anisotropies in intracellular transport},
  pdfauthor={Michał Balcerek; Adrian Pacheco-Pozo; Agnieszka Wyłomańska; Diego Krapf},
  pdfsubject={Scientific Reports manuscript},
  pdfkeywords={random walks, fractional Brownian motion, anisotropy, single-particle tracking, turning angles},
  bookmarksdepth=3
}

\newcommand{\R}{\mathbb{R}}

\newcommand{\dd}{\mathrm{d}}

\newcommand\abs[1]{\left|#1\right|}

\begin{document}

\title[Turning angle analysis reveals hidden anisotropies in intracellular transport]{Turning angle analysis reveals hidden anisotropies in intracellular transport}

\author*[1]{\fnm{Micha{\l}} \sur{Balcerek}}%
    \email{michal.balcerek@pwr.edu.pl}

\author[2]{\fnm{Adrian} \sur{Pacheco-Pozo}}
\author[1,3]{\fnm{Agnieszka} \sur{Wy{\l}omańska}}
\author[2]{\fnm{Diego} \sur{Krapf}}

\affil*[1]{\orgdiv{Faculty of Pure and Applied Mathematics, Hugo Steinhaus Center}, \orgname{Wroc{\l}aw University of Science and Technology}, \orgaddress{\postcode{50-370}, \city{Wroc{\l}aw}, \country{Poland}}}

\affil[2]{\orgdiv{School of Biomedical and Chemical Engineering}, \orgname{Colorado State University}, \orgaddress{\city{Fort Collins}, \state{Colorado}, \postcode{80523}, \country{USA}}}

\affil[3]{\orgdiv{Center of Trustworthy AI for Life Sciences - International Research Agendas Programme}, \orgname{University of Warsaw}, \orgaddress{\postcode{02-097}, \city{Warsaw}, \country{Poland}}}


\abstract{
    Molecular motion within living cells provides a window into the physical principles underlying intracellular organization and transport. Yet a fundamental limitation remains: measured trajectories are often short, noisy, and randomly oriented, rendering spatial anisotropies inaccessible to conventional analyses. We show that turning-angle statistics provide a robust approach for uncovering anisotropies in anomalous diffusion. Using a two-dimensional anisotropic fractional Brownian motion model with random orientations, we show theoretically and through simulations that turning-angle distributions preserve signatures of anisotropy. We apply this approach to single-particle tracking data, including quantum dots in the cytoplasm of HeLa cells and membrane proteins in hippocampal neurons. Turning angles reveal hidden anisotropies in the motion of quantum dots and K$_\textrm{v}$1.4 channels, and in specific dynamical states of glycoprotein CD4. These results establish turning-angle analysis as a powerful strategy for detecting organization in complex environments and reveal that anisotropic anomalous diffusion is an {often} overlooked feature of intracellular dynamics.
}

\keywords{random walks, FBM, isotropy, single-molecule tracking, directional changes}

\maketitle
\section{Introduction}
The dynamics of molecules are fundamental to the function of living cells and organisms. A primary experimental approach for studying these dynamics is single-particle tracking, in which individual fluorescently labeled molecules are localized with nanometer precision  \cite{manzo2015review,shen2017single}. Accordingly, the resulting images of individual particles are obtained in two dimensions using a camera, typically representing a two-dimensional projection of the three-dimensional cellular environment. Particle trajectories are then reconstructed by linking localizations across consecutive frames.  
The reconstructed trajectories provide a direct means of quantifying molecular motion and identifying the physical processes that govern subcellular organization. Single-particle tracking has therefore played a central role in studies of subcellular dynamic organization for more than two decades \cite{saxton,norregaard2017manipulation}. In particular, it is found that protein motion can deviate from normal diffusion, exhibiting marked differences from Einstein's laws of diffusion, i.e., from Brownian motion \cite{einstein05}. Normal diffusion yields a linear growth of the mean square displacement (MSD) $\langle X^2(t)\rangle \sim t$ as a function of time, where the brackets stand for averaging over different realizations of the measured trajectories of $X(t)$. However, the dynamics in complex cell environments often yield anomalous diffusion with non-linear MSD, which is often observed to follow a power law $\langle X^2(t)\rangle = 2D t^{\alpha}$, in which case $\alpha$ is the anomalous exponent and $D$ is the generalized diffusion coefficient with units of cm$^2$/s$^\alpha$ \cite{metzler2014,krapf2015mechanisms,hoefling}.  The motion is subdiffusive when $0<\alpha<1$, superdiffusive when $1<\alpha<2$, and normal diffusion is recovered for $\alpha=1$. 
Such deviations from normal behavior can be attributed to several physical mechanisms, such as viscoelastic environments \cite{tolic2004anomalous,weber2010bacterial}, crowding \cite{banks2005anomalous}, transient immobilization due to interactions \cite{torreno2016uncovering,mosqueira2020antibody}, bulk-mediated diffusion \cite{krapf2016strange}, and spatial heterogeneities \cite{han2020deciphering,waigh2023heterogeneous}. 
Given that each physical mechanism yields unique types of motion, a vast array of mathematical models has been developed to describe anomalous diffusion, including fractional Brownian motion (FBM)\cite{NessMandelbrot, kepten2011ergodicity,weiss2012}
, continuous time random walk (CTRW) \cite{montroll1969random,bib:Scher}, fractional Langevin equation \cite{deng2009ergodic,goychuk2009viscoelastic,lim2025anomalous}, heterogeneous diffusion \cite{cherstvy2013anomalous}, and subordinated diffusion processes \cite{dybiec2010subordinated,tabetal13,fox2021aging}. Identifying the stochastic model that governs anomalous diffusion provides critical insight into the physical mechanisms underlying experimental observations. 

Among the anomalous diffusion models, FBM is one of the most widely used. It is a Gaussian process with enough versatility to be applied to any anomalous exponent $0<\alpha<2$, i.e., both subdiffusion and superdiffusion.
FBM has been employed to model subcellular dynamics in the cytoplasm \cite{rienzo,sabri2020elucidating}, the nucleus \cite{kepten2011ergodicity,stadler2017non}, and the cell membrane \cite{krapf2015mechanisms}, as well as beads in viscoelastic media \cite{dolgushev2025evidence} and tracers in crowded fluids \cite{szymanski2009elucidating}. Given that the dynamics of particles in experimental data typically involve more than one dimension, the motion of tracers is often modeled with a multidimensional FBM. Classically, multidimensional FBM has been defined as a superposition of independent FBMs \cite{jeon2010fractional,krapf2019spectral,regnier2025visitation}. That is, a trajectory ${\bf R}(t) = (X(t),Y(t))$ assumes that $X(t)$ and $Y(t)$ are independent. In spite of the simplicity warranted by this approach, it can be hypothesized that motion within intracellular environments exhibits a preferred orientation and, thus, the FBM components may be correlated. For example, directional fibers such as microtubules and actin filaments could generate anisotropic motion, facilitating transport along their axes while hindering movement perpendicular to them \cite{sadegh2017plasma,krapf2018Compartmentalization,teves2020sperm}. Similar situations can be introduced by the sheets and branched tubules characteristic of the endoplasmic reticulum morphology \cite{nixon2016increased}, as well as the complex, tortuous, and highly connected meshwork that forms extracellular spaces \cite{sykova2008diffusion}. Such anisotropy naturally introduces correlations in the observed components of the molecule trajectory unless the two components are purposely chosen to be aligned with the direction of the underlying structure.

From the point of view of data analysis, it is possible to characterize spatial dependencies by computing the cross-covariance function or the cross power spectral density \cite{balcerek2026two}. However, a limitation of these approaches is that they rely on the environment being homogeneous in the sense that all trajectories should have the same directional preference.
Yet, in the cell environment, the preferred direction can be heterogeneous. For example, within a single cell, it is possible that the fibers in distinct regions are aligned in completely different directions. Moreover, when cells plated on glass are imaged, each cell attaches in a random orientation, and therefore, the directionality of trajectories differs from cell to cell. Thus, unraveling anisotropies and spatial correlations in the dynamics is not possible at the ensemble level using standard tools. In particular, correlations captured by cross-power spectral density or cross-covariance functions can average out when the trajectories are embedded in randomly oriented frames of reference. Alternatively, it is possible to infer anisotropies in single long trajectories from metrics such as the trajectory eccentricity \cite{malacrida2017visualization} or by estimating the non-diagonal terms of the two-dimensional diffusion tensor \cite{alexander2007diffusion, di2016diffusion}. However, in many experimental situations, one must deal with short trajectories. The situation is further complicated when individual trajectories exhibit motion changes, requiring the analysis to be performed on short trajectory segments \cite{munoz2023quantitative}.
As a consequence, an alternative approach that is invariant to rotational alignment and capable of capturing anisotropies at the ensemble level is needed. One potential method that can reveal anisotropies and spatial correlations is the distribution of turning angles, also referred to as the distribution of directional change \cite{burov2013distribution,sadegh2017plasma}, which is intrinsically invariant under rotations and therefore independent of the reference frame orientation. 

In this article, we employ turning angle analysis in combination with multidimensional FBM models to assess whether trajectories of subdiffusive particles in live mammalian cells exhibit spatial anisotropies. As predicted, we observe that traditional methods, such as covariance matrices, cannot give an answer regarding correlations between components when the trajectories are randomly rotated.
Instead, the analysis of the turning angle distribution shows that different intracellular dynamics exhibit previously unrecognized anisotropies. Specifically, we  analyze the motion of semiconductor nanocrystals that were inserted into the cytoplasm of live HeLa cells and the motion of two different membrane proteins in the soma of live hippocampal neurons: the voltage-gated potassium channels K$_\textrm{v}$1.4 and the glycoprotein CD4. The degree of anisotropy among these particles differs substantially.
We validate the turning angle analysis using numerical simulations with and without localization noise under a broad range of parameters, considering subdiffusive, normally diffusive, and superdiffusive modes of motion. 
This method provides a sensitive means of detecting hidden spatial anisotropies in intracellular dynamics, particularly in systems with randomly oriented trajectories where conventional approaches fail. This information can improve our understanding of how the cellular environment and molecular interactions constrain motion at the nanoscale, enabling a deeper and more accurate characterization of molecular dynamics in complex environments.

\section{Results}

\subsection{Anisotropic two-dimensional FBM (2D-FBM) with random rotations}
We focus on two-dimensional trajectories, as they represent the most common type of data collected from single-particle tracking. We begin by describing the motion of 2D-FBM with independent components. The process is constructed from two independent FBMs sharing the same Hurst exponent $0<H<1$. The anomalous diffusion exponent $\alpha$ is linked to the Hurst exponent $H$ through the relation $\alpha=2H$. 
The anisotropic process with independent components $\widehat{\mathbf{R}}(t)=(\widehat{R}_1(t),\widehat{R}_2(t))$ is therefore defined by the covariance structure
\begin{align}
\Big\langle \widehat{R}_i(t)\widehat{R}_j(s) \Big\rangle=D_i[i=j] \Big(t^{2H}+s^{2H}-|t-s|^{2H} \Big),
\end{align}
where the Iverson bracket $[i=j]$ is $1$ when $i=j$, and $0$ otherwise.
Without loss of generality,  $D_1\geq D_2>0$ are the generalized diffusion coefficients in the two intrinsic directions. We characterize the spatial anisotropy by the ratio
\begin{align}
r= D_2/D_1,\qquad 0<r\leq 1.
\end{align}
The case $r=1$ corresponds to isotropic motion, whereas $r<1$ indicates that displacements along one direction are more strongly constrained than along the other. 
When evaluating the increments over a time interval $\Delta$, 
$\delta_\Delta\widehat{\mathbf{R}}(t)=\widehat{\mathbf{R}}(t+\Delta)-\widehat{\mathbf{R}}(t)$,
the covariance matrix of the increments is
\begin{align}
\Big\langle \delta_\Delta\widehat{\mathbf{R}}(t+h) \delta_\Delta\widehat{\mathbf{R}}(t)^\mathsf{T}\Big\rangle=\Big(|h+\Delta|^{2H}+|h-\Delta|^{2H}-2|h|^{2H}\Big)
\begin{bmatrix} D_1 & 0 \\
    0 & D_2
\end{bmatrix}.
\end{align}

In experimental data, the orientation of the principal directions is generally unknown and may differ between trajectories. After being rotated by an angle $\varphi$, a trajectory with originally independent components becomes $\mathbf{R}(t)=\mathbf{A}_\varphi \, \widehat{\mathbf{R}}(t)$, where $\mathbf{A}_\varphi$ is the rotation matrix,
\begin{align}
\mathbf{A}_\varphi=
    \begin{bmatrix}
    \cos\varphi & -\sin\varphi\\
    \sin\varphi & \phantom{-}\cos\varphi
    \end{bmatrix}.
\end{align}
Although the intrinsic components are independent, the components of the rotated trajectory $\mathbf{R}(t)=(X(t),Y(t))$ are generally correlated. Their increment covariance matrix is
\begin{align}
\Big\langle \delta_\Delta\mathbf{R}(t+h) \delta_\Delta\mathbf{R}(t)^\mathsf{T}\Big\rangle = \Big(|h+\Delta|^{2H}+|h-\Delta|^{2H}-2|h|^{2H}\Big) \mathbf{A}_\varphi
\begin{bmatrix}
D_1 & 0 \\
0 & D_2
\end{bmatrix}
\mathbf{A}_\varphi^\mathsf{T}.
\end{align}
In particular, the non-diagonal terms, i.e., the increment cross-covariance function, are
\begin{align}
C_{xy}(h)\equiv \Big\langle \delta_\Delta X(t+h) \delta_\Delta Y(t) \Big\rangle & =(D_1-D_2)\sin\varphi\cos\varphi\,\Big(|h+\Delta|^{2H}+|h-\Delta|^{2H}-2|h|^{2H}\Big) \nonumber \\
& = D_1(1-r)\sin\varphi\cos\varphi\Big(|h+\Delta|^{2H}+|h-\Delta|^{2H}-2|h|^{2H}\Big)
\end{align}
Therefore, the cross-covariance function of the increments depends not only on the intrinsic anisotropy~$r$, but also on the orientation of the trajectory relative to the observation frame. It vanishes both when $r=1$, i.e., the motion is isotropic, and when the coordinate axes coincide with the principal directions. Furthermore, the average cross-covariance function of the increments also vanishes after averaging over uniformly distributed orientations.

\begin{figure}
    \centering
    \includegraphics[width=0.5\linewidth]{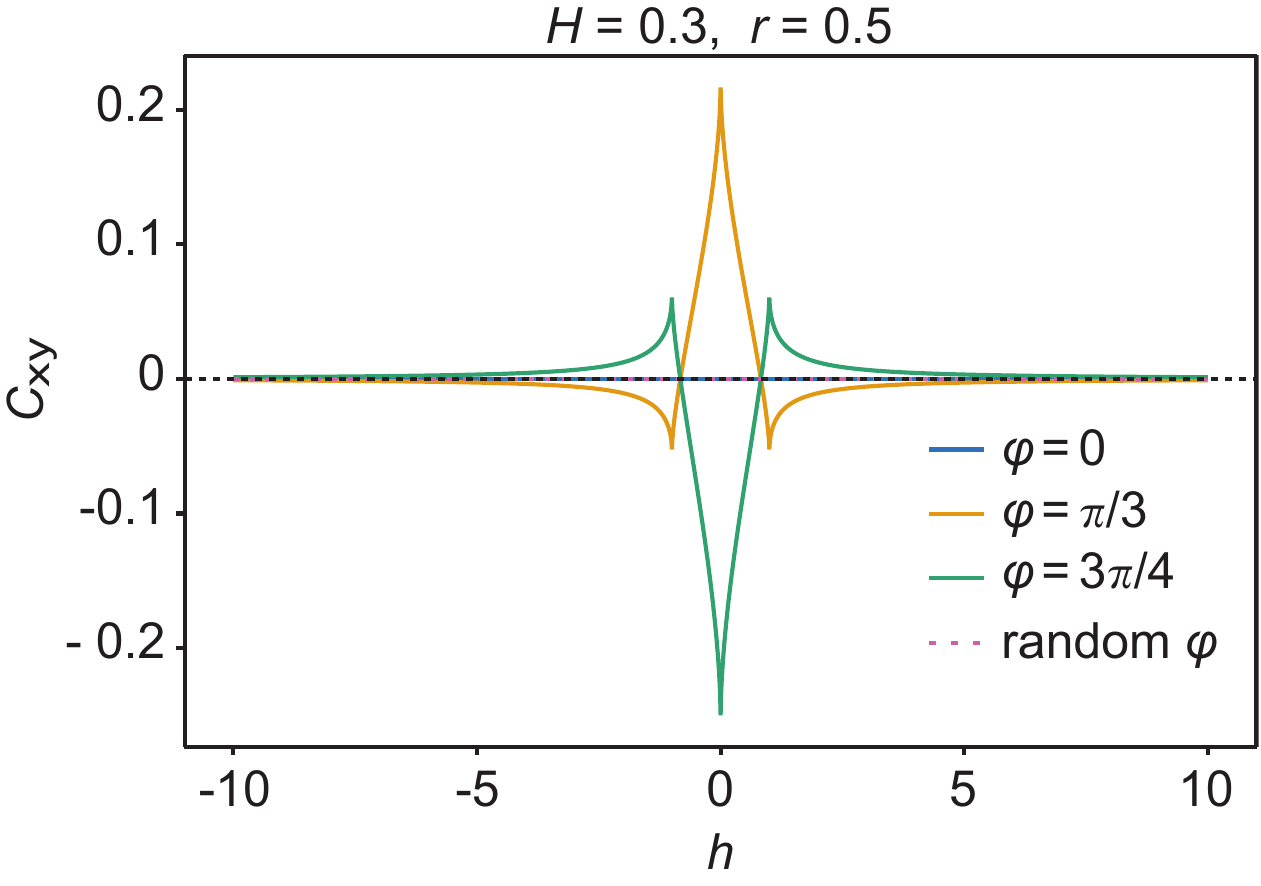}
   \caption{{\bf Cross-covariance functions $C_{xy}(h)$ of the increments of anisotropic 2D-FBM.} The process has a subdiffusive Hurst exponent $H=0.3$, anisotropy ratio $r=0.5$, and $D_1=0.5$. Four cases are shown: the process where the two components along $x$ and $y$ are independent {($\varphi=0$)}, after the trajectories are rotated by $\varphi=60^{\circ}$ and $\varphi=135^{\circ}$, and when each trajectory undergoes a random rotation with a uniformly distributed {orientation} (dotted line).}
    \label{fig:1_ccov}
\end{figure}

Figure~\ref{fig:1_ccov} illustrates the effect of rotations for subdiffusive FBM,  $\mathbf{R}(t)$, with $H=0.3$ and $r=0.5$. In the intrinsic frame, corresponding to $\varphi=0$, the two components are independent, and their cross-covariance function $C_{xy}(h)$ vanishes. Fixed rotations ($\varphi = \pi/3$ and $\varphi = 3\pi/4$) generate non-zero cross-covariance functions whose magnitude and sign depend on $\varphi$. For $H<1/2$, the increment covariance kernel, i.e., the time-dependent term $|h+\Delta|^{2H}+|h-\Delta|^{2H}-2|h|^{2H}$,  has a positive value at zero lag and negative values at neighboring lags, while the prefactor, $(D_1-D_2)\cos\varphi \sin\varphi$ is determined by the rotation angle and can either preserve or reverse this structure. When each trajectory is rotated by a random angle uniformly distributed between $0$ and $2\pi$ , the cross-covariance function averages to zero at all lag times.

As an alternative approach, one may consider spatially correlated multidimensional FBM to account for anisotropies \cite{lavancier2009covariance,amblard2010basic,coeurjolly2010multivariate}. 
In our previous article \cite{balcerek2026two}, we proposed a construction of 2D-FBM as a stochastic process $\mathbf{R}(t)=(X(t),Y(t))$, where $X(t)$ and $Y(t)$ were defined as FBMs with generalized diffusion coefficients $D_x$ and $D_y$, and correlation coefficient $\rho$ between them. Such a model, for which Brownian motion with correlated components constitutes a special case \cite{balcerek2025two}, is equivalent to the one presented here. However, the parameters $D_x$, $D_y$, and $\rho$ depend on the orientation of the frame of reference. After rotating the diffusion tensor to its principal axes, the two components become independent, while $D_1$ and $D_2$ correspond to the eigenvalues of the diffusion tensor. In other words, the two components become independent when the observation axes coincide with the intrinsic principal directions of the motion. Consequently, $r=D_2/D_1$ provides a rotation-invariant measure of the intrinsic anisotropy. The relations between the two different parametrizations are presented in the Supplementary Methods \ref{si:b2}.

\subsection{Turning angle analysis}
\begin{figure}
    \centering
    \includegraphics[width=0.9\linewidth]{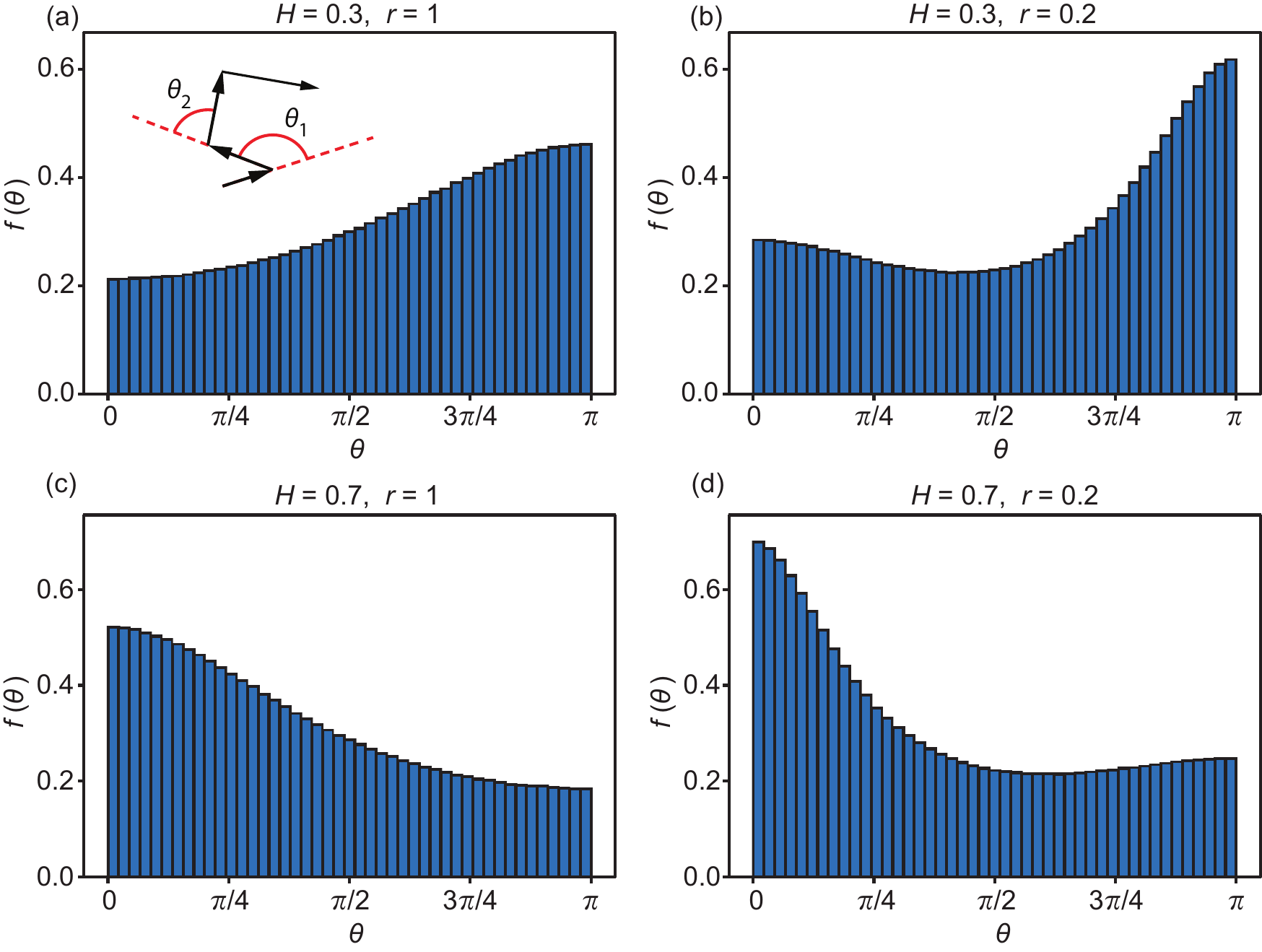}
    \caption{{\bf Distribution of turning angles in simulated 2D-FBM trajectories.} (a) Isotropic subdiffusive FBM, $H=0.3, r=1$. The inset illustrates the definition of turning angles $\theta$. (b) Anisotropic subdiffusive FBM, $H=0.3, r=0.2$. (c) Isotropic superdiffusive FBM, $H=0.7, r=1$. (d)~Anisotropic superdiffusive FBM, $H=0.7, r=0.2$.}
    \label{fig:2_TurnAngles}
\end{figure}

To elucidate anisotropies in systems that can be affected by random rotations, such as a live cell, it is critical to employ statistics that are invariant under rotations. One such metric is the angle between subsequent increments, i.e., the turning angles $\theta$, as depicted in the inset of Fig.~\ref{fig:2_TurnAngles}a. The distribution of turning angles provides a robust metric for capturing intrinsic geometric properties of motion, even in the presence of random rotations, while remaining straightforward to compute from trajectory data (see Supplementary Information). From a physical perspective, the turning angles capture local directional changes of motion, rather than absolute position.
In a simple Brownian motion (i.e., 2D-FBM with $H=1/2$) with independent components, the turning angles are uniformly distributed. Contrastingly, when increments at different times are correlated \cite{burov2013distribution,sadegh2017plasma} or when the random walk is not isotropic  \cite{tierno2016enhanced,wadkin2018correlated}, the distribution of turning angles is not uniform. Supplementary Fig. \ref{fig:S1} shows the distribution of turning angles obtained from numerical simulations of two-dimensional Brownian motion. As expected \cite{balcerek2025two}, the turning angles are uniformly distributed when the Brownian motion is isotropic, whereas deviations from the uniform distribution are observed when anisotropies are introduced.

Figure~\ref{fig:2_TurnAngles} shows distributions of turning angles obtained from simulations of subdiffusive ($H=0.3$) and superdiffusive ($H=0.7$) 2D-FBM. A prevalent feature of FBM is a non-symmetric distribution of turning angles. When the FBM is subdiffusive ($H<1/2$), the tracer is more likely to turn back than to move forward, i.e., it is antipersistent. As a consequence, the distribution has a maximum at a turning angle $\theta=\pi$. In contrast, superdiffusive FBM ($H>1/2$) is persistent, that is, the tracer is more likely
to keep moving in the same direction rather than turning back, and consequently the maximum of the distribution is found at a turning angle $\theta=0$. Moreover, the distribution of turning angles is influenced not only by the Hurst exponent but also by the presence of anisotropies. For isotropic subdiffusive 2D-FBM ($H<1/2$, $r = 1$), the turning angle distribution attains its minimum at $\theta = 0$, indicating that consecutive increments are least likely to align in the same direction. For isotropic superdiffusive FBM ($H>1/2$, $r = 1$), the minimum lies at $\theta=\pi$. However, when $r \neq 1$, i.e., the random walk is inherently anisotropic, the minimum of the distribution can occur at an intermediate angle between $0$ and $\pi$.

\subsection{Experimental trajectories in live cells}

\begin{figure}
    \centering
    \includegraphics[width=1\linewidth]{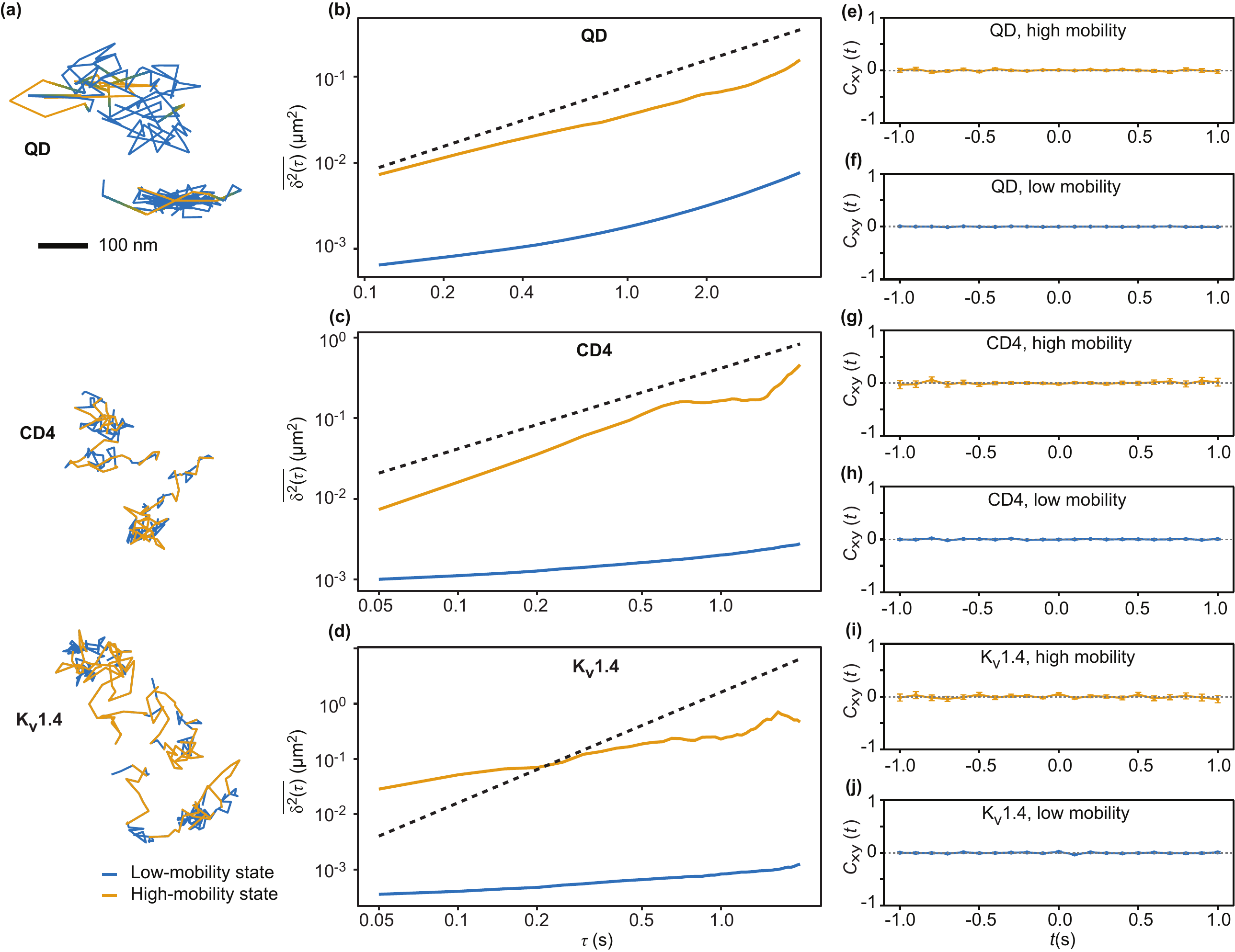}
    \caption{{\bf Experimental trajectories in live cells.} (a) Sample trajectories of quantum dots in the cytoplasm of HeLa cells, CD4 receptors, and K$_\textrm{v}$1.4 channels on the membrane of hippocampal neurons. Low- and high-mobility states identified using the local convex-hull method are shown in blue and yellow, respectively. (b--d) TA-MSD for the corresponding states and datasets. The dashed lines indicate linear scaling with lag time. (e--j) Ensemble cross-covariance functions of normalized increments for the high- and low-mobility states of the three experimental systems.}
    \label{fig:3_Experiments}
\end{figure}

We evaluated the dynamics in living cells of three systems that have been thoroughly characterized previously. Namely, we analyzed the motion of two transmembrane proteins on the surface of hippocampal neurons and that of quantum dots introduced into the cytoplasm of HeLa cells. For membrane proteins, we studied the voltage-gated potassium channel K$_\textrm{v}$1.4-tagged with an extracellular CF640R fluorophore via biotin-streptavidin and CD4 receptors tagged with a CF640R-conjugated antibody \cite{fox2013plasma,sikora_diego,weron2017ergodicity}. Although the dynamics of these membrane proteins have been characterized using a diverse set of classical metrics, these statistics are not capable of testing whether the trajectories are influenced by underlying anisotropies. These dynamics have been previously evaluated in terms of their ensemble-averaged and time-averaged MSD (TA-MSD) \cite{weron2017ergodicity}, dynamical functionals \cite{weron2017ergodicity}, fractional time-series models \cite{burnecki2019identifying}, and recurrence analysis \cite{sikora_diego}. One key aspect that these analyses have revealed is that these proteins alternate between two states, with one of the states exploring smaller membrane areas. One of these states was found to have a heavy-tailed distribution of dwell times, inducing, in turn, weak ergodicity breaking \cite{weron2017ergodicity, weietal11}.  Figure~\ref{fig:3_Experiments}a shows representative trajectories, with the two states shown in different colors. 

In addition to the motion on the cell surface, we also analyzed the dynamics of quantum dots introduced by bead loading into the cytoplasm of HeLa cells \cite{sabri2020elucidating}. While these particles move in three dimensions, the imaging is performed only in two of them, capturing two components of the random walk. As was the case for membrane proteins, the quantum dot dynamics have been analyzed with a large set of metrics including MSD \cite{sabri2020elucidating}, autocorrelation functions \cite{sabri2020elucidating}, power spectral density \cite{sposini2022towards}, hidden Markov models \cite{janczura2021identifying},  increment nonstationarity \cite{vilk2022unravelling}, intermediate scattering function (ISF) \cite{dieball2022scattering}, and dynamic structure factor (DSF) \cite{dieball2022scattering}. The motion of quantum dots also alternates between two different diffusive states, but in this case, the dynamics are ergodic because these states have exponentially distributed dwell times \cite{sabri2020elucidating,janczura2021identifying}. Representative quantum dot trajectories are shown in Fig.~\ref{fig:3_Experiments}a. These states have been found to be accurately described by FBM, with the transitions being represented by changes in the generalized diffusion coefficient and/or the anomalous exponent \cite{sabri2020elucidating,balcerek2023modelling}.

Given that it has been established the analyzed trajectories alternate between two different modes of motion, we first segmented the trajectories into two different states. For this purpose, we used the local convex hull to obtain states of low and high mobility  \cite{sabri2020elucidating, seg1}. {In the current work, we rely on the previously established number of dynamical classes in these experimental trajectories and do not attempt to re-evaluate the number of states.} The Hurst exponent of each of these states was found using the TA-MSD, 
\begin{align}
    \overline{\delta^2(\tau)} = \frac{1}{T-\tau}\int_0^{T-\tau} |{\bf R}(t+\tau)-{\bf R}(t)|^2 \dd t,
\end{align}
where  $\tau$ is the lag time. Namely, we obtained $\alpha$ (and $H=\alpha/2$) from the relation $\overline{\delta^2(\tau)}\sim\tau^{\alpha}$.
Figs.~\ref{fig:3_Experiments}b-d show the TA-MSD of the two states for the three systems under consideration. The anomalous exponent $\alpha$ of each state was estimated via linear regression of the first five data points of the TA-MSD in log–log space, and the Hurst exponent was found. For the high mobility (h) and low mobility (l) states, $H_{\rm QD,h}=0.27$, $H_{\rm QD,l}=0.19$,  $H_{\rm CD4,h}=0.37$, $H_{\rm CD4,l}=0.12$, $H_{\rm Kv,h}=0.14$, $H_{\rm Kv,l}=0.06$. The cross-covariance of the increments for each state is presented in Figs.~\ref{fig:3_Experiments}e-j, which was found to vanish in all cases. The zero cross-covariance is consistent with a system where the trajectories are randomly rotated, as expected for cells plated on glass. Thus, the cross-covariance cannot be used to evaluate correlations among components here.

\begin{figure}
    \centering
    \includegraphics[width=1\linewidth]{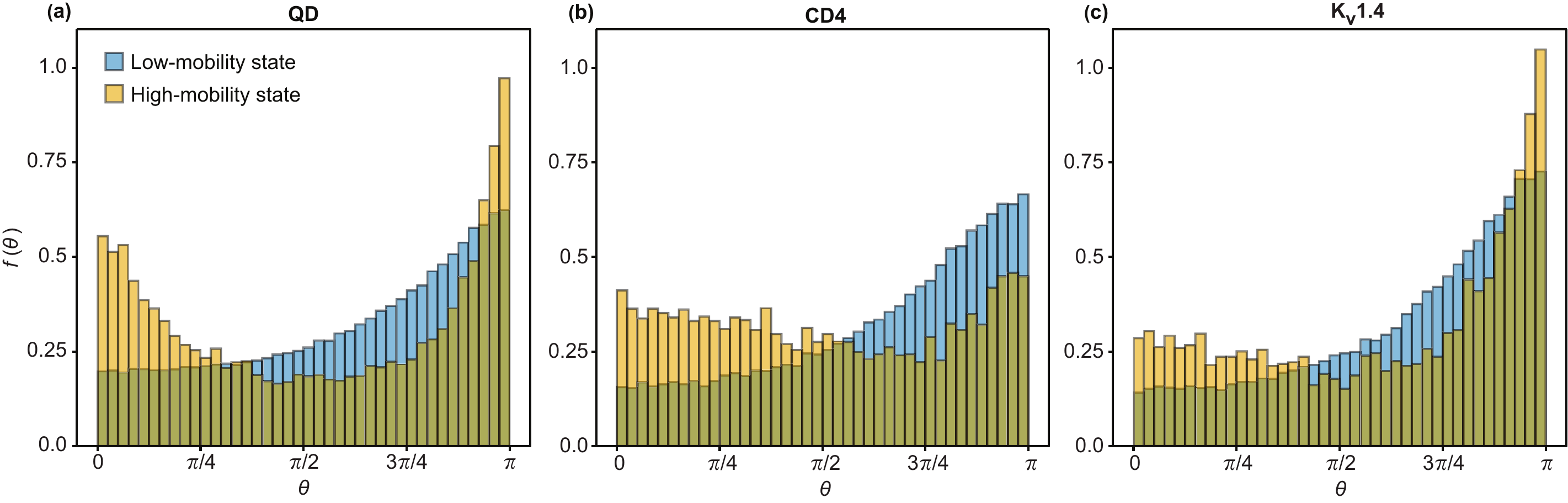}
    \caption{{\bf Distributions of turning angle in experimental data.} Probability density functions of the turning angles for (a) quantum dots, (b) CD4 receptors, and (c) K$_\textrm{v}$1.4 channels. The distributions are shown separately for the low- and high-mobility states.}
    \label{fig:4_Experiments_phi}
\end{figure}

To assess whether the components are isotropic, we evaluated the distribution of turning angles. 
The distributions for the two states in the three systems exhibit interesting, non-trivial features. These distributions are shown in Fig.~\ref{fig:4_Experiments_phi}. In particular, in most distributions the mode (maximum) is localized at $\theta=\pi$ as predicted for subdiffusive, antipersistent types of motion. Furthermore, in several cases, the minimum of the distribution is not at $\theta=0$ but somewhere within the interval $0<\theta<\pi$, a fingerprint of the motion being anisotropic. This analysis provides evidence for the need for anisotropic multidimensional random walks in the characterization of membrane proteins and in the motion of particles in the cytoplasm.

%

\begin{figure}
    \centering
    \includegraphics[width=1\linewidth]{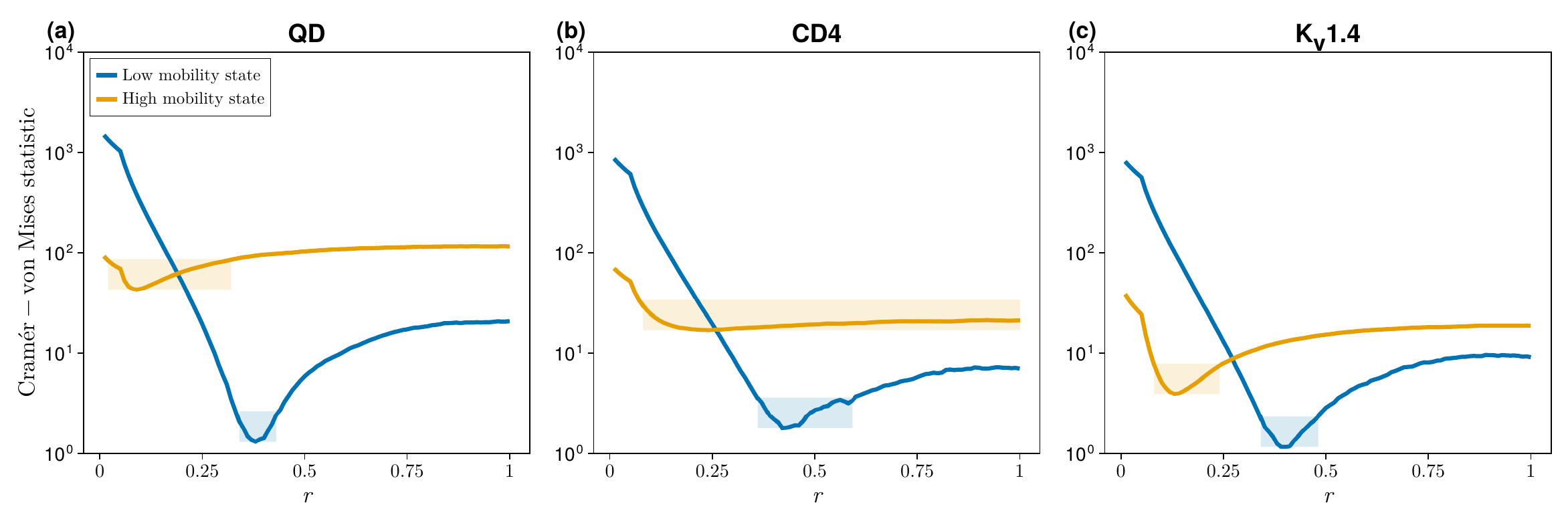}
    \caption{{\bf Cram\'er--von Mises distance between the experimental turning-angle distributions and the corresponding simulated look-up tables.} The distance is presented as a function of the anisotropy ratio $r=D_2/D_1$ for (a) quantum dots, (b) CD4 receptors, and (c) K$_\textrm{v}$1.4 channels. Results are shown for the low-mobility and high-mobility states. The shaded regions indicate the uncertainty ranges defined by values of $r$ for which the Cram\'er--von Mises statistic does not exceed twice its minimum value.}
    \label{fig:5_Experiments_rho}
\end{figure}

To estimate the anisotropy parameter $r$, we built look-up tables for different anisotropy parameters and a specific Hurst exponent, and compared the measured distribution from experimental data to those of the look-up tables (see Supplementary Methods \ref{si:b4} for details). Specifically, we built look-up tables using simulated trajectories with the estimated Hurst exponent $H$ of each state (Supplementary Fig. \ref{fig:S2}). Finally, we identified the value of $r$ that minimizes the discrepancy between the distributions of turning angles values using the Cram\'er--von Mises (CvM) distance \cite{anderson1962distribution}. Figure~\ref{fig:5_Experiments_rho} shows the CvM distances of the measured turning angle distributions to the distributions in the look-up tables. The $r$ value that yields the minimum distance corresponds to the estimated parameter. Furthermore, we quantified the uncertainty in the estimated anisotropy ratio by determining the range of $r$ that yields CvM distances not exceeding twice the minimum. The uncertainty ranges for $r$ are shown in Fig.~\ref{fig:5_Experiments_rho} as shaded regions.

Clear anisotropies are found in both mobility states of K$_\textrm{v}$1.4 channels and quantum dots as well as in the low mobility state of CD4 receptors. The motion of the quantum dots in the cytoplasm yields $r_{\rm QD,h}=  0.17\pm 0.15$ and $r_{\rm QD,l}= 0.38 \pm 0.04$ for the high- and low-mobility states, respectively. 
The K$_\textrm{v}$1.4 channels have $r_{\rm Kv,h}= 0.16\pm 0.08$ and $r_{\rm Kv,l}= 0.41\pm 0.07$. While anisotropy is found in the low-mobility state of CD4 receptors, the test of anisotropy was inconclusive for this molecule in the high-mobility state. For CD4, the low mobility state had $r_{\rm CD4,l}=  0.48\pm 0.12$, but the high-mobility state displayed $r_{\rm CD4,h}= 0.54\pm 0.46$.  Thus, we cannot reject the hypothesis that the high-mobility motion of CD4 receptors is isotropic. We also considered alternative statistics to identify the discrepancy between the measured turning angle distributions and the distributions in the look-up table, including $L^1, L^2, L^\infty$, Kolmogorov--Smirnov, and Wasserstein. However, the results were robust independent of the distance employed (Supplementary Fig. \ref{fig:S4}). {To compare the turning angle distribution obtained experimentally with that predicted by simulations reproducing the observed anisotropy, Supplementary Fig. \ref{fig:S5} shows the distribution for the high mobility state of K$_\textrm{v}$1.4 together with results from simulations both within and just outside the uncertainty region of the estimated anisotropy. Specifically, we show the distributions corresponding to $r=0.13$ and $r=0.4$. Visual comparison with the experimental data demonstrates that the method is robust to modest variations in the anisotropy parameter.}


\subsection{Effect of localization noise on anisotropy estimation}

The anisotropy parameter estimates for both states of K$_\textrm{v}$1.4 in the neuronal soma and low-mobility state of quantum dots in the cytoplasm are highly accurate, exhibiting narrow uncertainty range with widths smaller than $0.2$. However, this is not always the case. For example, the high-mobility state of quantum dots has an anisotropy uncertainty region of substantial width, $\Delta r=0.3$. Furthermore, the uncertainty region of CD4 receptors is so large ($\Delta r = 0.92$) that it is impossible to assess whether this state has an anisotropic mode of motion.
The uncertainty in the estimation of anisotropy can be traced to different mechanisms. One of them is the localization error, which is ubiquitous in single-particle tracking \cite{savin2005static,yildiz2005fluorescence}. Localization error arises from the finite number of photons collected during the exposure time from the emission of an individual fluorophore. These photons arrive at random locations within the point spread function in the imaging camera; therefore, the localization of the emitter is limited by the central limit theorem. Mathematically, this type of error is considered as additive Gaussian white noise to the true trajectory \cite{sposini2022towards}.   

\begin{figure}
    \centering
    \includegraphics[width=1\linewidth]{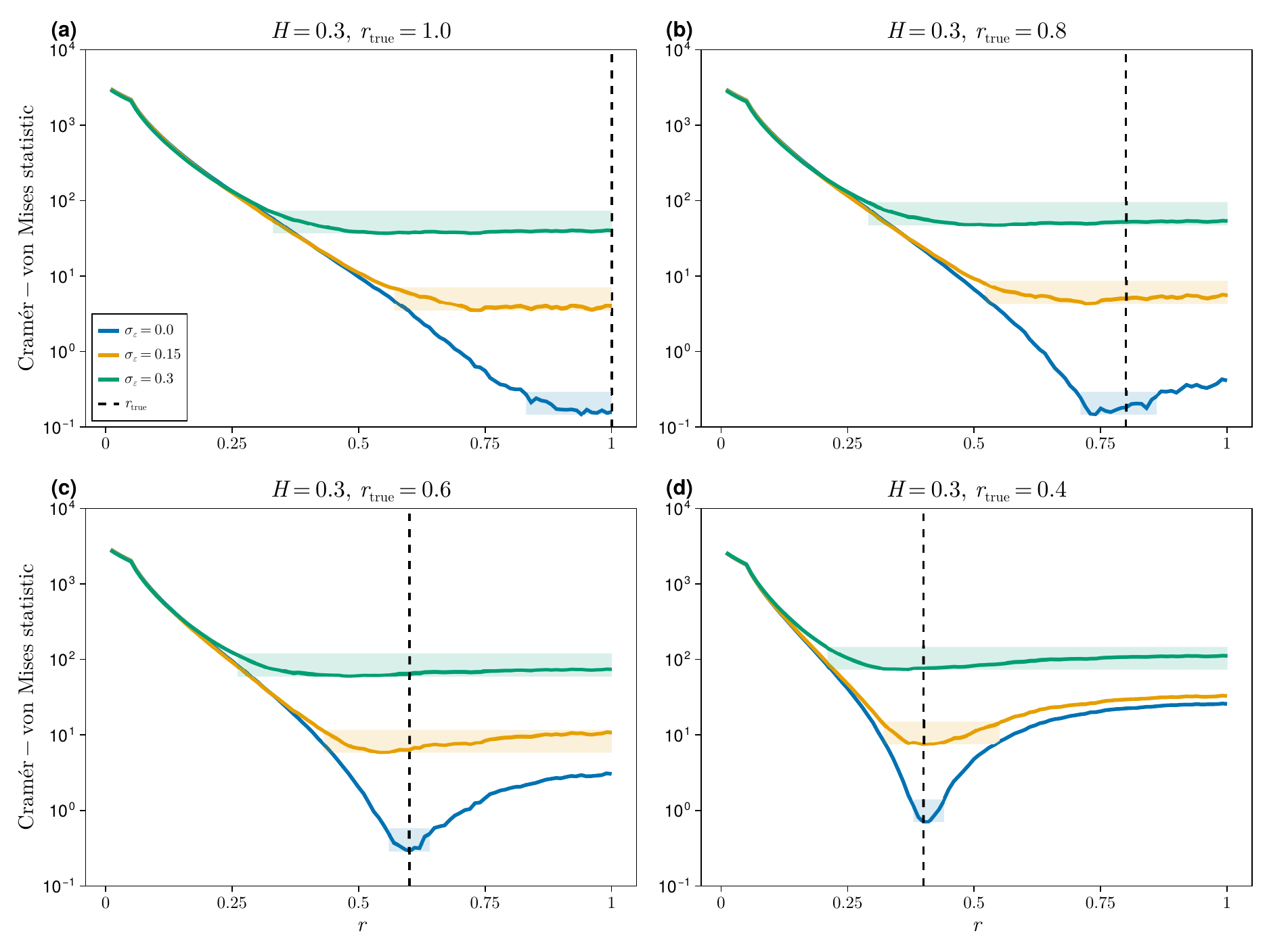}
    \caption{{\bf Cram\'er--von Mises statistic as a function of the anisotropy ratio $r$ for simulated anisotropic 2D-FBM trajectories.} The realizations have $H=0.3$ and true anisotropy ratios (a) $r_{\mathrm{true}}=1.0$, (b) $r_{\mathrm{true}}=0.8$, (c) $r_{\mathrm{true}}=0.6$, and (d) $r_{\mathrm{true}}=0.4$. Curves correspond to additive Gaussian localization noise with standard deviations $\sigma_\varepsilon=0$, $0.15$, and $0.30$. The vertical dashed lines indicate the true values $r_{\mathrm{true}}$. The shaded regions show the uncertainty ranges defined by Cram\'er--von Mises values not exceeding twice the corresponding minimum.}
    \label{fig:6_rho-sub}
\end{figure}

We assessed the impact of localization error on the estimation of anisotropy using simulations of 2D-FBM in the presence of additive Gaussian noise. In the same manner as performed for the experimental trajectories, the distributions of turning angles in the realizations with added noise were compared to a look-up table for the noiseless process, and the uncertainty range for the anisotropy parameters were determined. Figure~\ref{fig:6_rho-sub} shows the effect of the added noise in the estimation of the anisotropy for parameters $r$ between 0.4 and 1 and Hurst exponent $H=0.3$, i.e., subdiffusive FBM. The trajectories were normalized so that the higher-diffusivity principal component was unity, and two levels of noise with standard deviations of $0.15$ and $0.30$ were added to the trajectories. It was found that in the simulations without noise, the uncertainty ranges of the estimated anisotropy parameter are relatively narrow, but as the noise increases, the width of the uncertainty region increases. A similar effect was observed for anisotropic Brownian motion ($H=1/2$, Supplementary Fig. \ref{fig:S3}) and for superdiffusive FBM ($H>1/2$, Supplementary Fig. \ref{fig:S6}). 

\section{Discussion and Conclusions}

The analysis of single-particle tracking data can reveal fundamental biological insights. Single-particle tracking is used to probe local properties of cells and tissues, and it provides important information about interactions between cell components as well as dynamic interactions with the local environment. Besides intracellular dynamics, similar tools are employed in the analysis of diverse systems such as the paths of ants and migrating birds \cite{vilk2022unravelling}.  One important aspect of the dynamics is elucidating anisotropies in the trajectories. However, accurate estimation of anisotropy is challenging due to several aspects. First, molecules within live cells are often found to alternate between different states of motion. In particular, dwell times in each state can be short, adding challenges from a statistical perspective. Second, while individual trajectories can be locally anisotropic, different trajectories or segments within a trajectory can be oriented along different axes, posing significant complexities and restricting the use of ensemble averages. Third, experimental trajectories are inherently affected by noise that corrupts the data.   

We employed the distribution of turning angles to reveal trajectory anisotropies in two-dimensional FBM. This metric has the advantage that it is not affected by the orientation of a given trajectory and can thus be averaged over an ensemble, even when individual trajectories are randomly rotated. In terms of data analysis, incorporating turning angle distributions shifts the focus from second-order statistics, such as covariance, to directional statistics, enabling a better distinction between different modes of motion and accounting for anisotropic effects during model fitting and interpretation of results. This approach can lead to more accurate parameter estimation and the appropriate selection of physical models that describe particle transport.
In general, turning angle analysis is a powerful and robust complementary tool that uncovers otherwise hidden anisotropies in complex biological systems.
We found anisotropic motion both in the dynamics of particles in the cytoplasm and molecules on the plasma membrane. {Our results are consistent with previous studies of intracellular anisotropies in systems where the preferred direction of motion could be independently determined. For example, imaging of 500-nm endocytosed polystyrene beads in cells cultured on micropatterned strips revealed that cell aspect ratio alters cytoskeletal organization and induces anisotropic motion \cite{gupta2019anisotropic}. In that system, the direction of anisotropy is known a priori because the cells are geometrically confined along a single axis. Guided anisotropic diffusion has been observed for the CCCTC-binding factor in the nucleus, which appears to originate from transient trapping in spatially localized regions biasing its motion along preferred directions \cite{hansen2020guided}. Recently, peroxisomes, membrane-bound cytoplasmic organelles with diameters of 0.1-1.0 $\mu$m, were found to exhibit significant anisotropic subdiffusion \cite{sarkar2026anisotropic}. In that system, the abundance of fluorescently labeled peroxisomes and their long track lengths provided a very large number of trajectories, making it possible to determine the local preferred direction of motion directly from the data. Altogether,} these results suggest that anisotropies may be a widespread feature in live cells. As a consequence, we infer that the characterization of trajectories within live cells should not assume a priori that the motion is isotropic. {In particular, the common way of estimated motion parameters using a single diffusivity measure may be misleading.}

Isotropy was assessed using the anisotropy parameter. We found that the uncertainty in the estimation of the anisotropy parameter depends on the experimental noise. In particular, we evaluated the estimation uncertainty as a function of static localization noise and found that greater noise leads to greater uncertainty. While only static noise was explicitly evaluated in this work, other sources of noise can also reduce the estimation precision. In particular, single-particle tracking data is affected by dynamic error due to finite exposure time \cite{savin2005static,sposini2022towards}.
Here we focus on 2D-FBM cases. While FBM is widespread in biophysics, it is not universal. Cases with multiple states, such as switching FBM \cite{balcerek2023modelling}, can be directly analyzed with the described methodology by segmenting the trajectories and analyzing each state independently, as performed in this work. Furthermore, extending the work presented here to other models \cite{andi21} is possible. Proper estimation of anisotropy only requires simulations of these processes to obtain look-up tables of the distribution of turning angles.

The results presented here demonstrate that analyzing turning angles provides novel insights into intracellular phenomena. In particular, it reveals anisotropies that remained hidden under traditional evaluations. {The described method is particularly useful when the preferred direction is unknown, and it can be used to infer the degree of anisotropy even for sparse short trajectories. Once anisotropy is found, the specific direction of motion may be independently investigated, for example by independently measuring the alignment of intracellular structures.} Although this methodology is demonstrated within a specific model, fractional Brownian motion in live cells, we anticipate its utility across a broad spectrum of applications where inter-component dependencies in multidimensional models are inherently present. {Beyond differences in diffusivity, other mechanisms may also give rise to motion anisotropy. For example, anisotropic motion can result from confinement that restricts displacement along one direction, or from transient tethering of particles to filaments that permit movement primarily along their axis. Understanding how these and other mechanisms shape turning-angle distributions will require further theoretical and experimental investigation.}

\section{Methods}

\subsection{Membrane proteins in hippocampal neurons}
K$_\textrm{v}$1.4 channels and CD4 receptors were imaged as previously described \cite{fox2013plasma,sikora_diego,weron2017ergodicity}. Briefly, rat hippocampal neurons were cultured and imaged on glass-bottom 35-mm dishes with No. 1.5 coverslips (MatTek) coated with poly-L-lysine. Animals were used according to protocols approved by the Institutional Animal Care and Use Committee of Colorado State University (Animal Welfare Assurance Number A3572-01). Dissections were performed after anesthesia with isoflurane followed by decapitation. Hippocampal tissue was dissected from the brains of E18 embryos and neurons were plated on coated glass-bottom dishes.

K$_\textrm{v}$1.4 constructs were modified to contain an extracellular biotin acceptor domain in an extracellular loop. Neuronal transfections were performed after 4–6 days in vitro in culture using Lipofectamine 2000 (LifeTechnologies). Bacterial biotin ligase pSec-BirA was cotransfected to biotinylate the channel.
Labelling of the channel was performed before imaging by incubating for 10 min with streptavidin-conjugated CF640R (Biotium) at $37 ^{\circ}$C in the presence of bovine serum albumin (cat. A0281, Sigma).
For experiments with CD4, cells were incubated for 10 min at $37 ^{\circ}$C with a monoclonal antibody against CD4 (MABF573, Millipore) conjugated with a CF640R fluorophore (antibody labelling kit Mix-n-stain CF640R, Biotium).
Total internal reflection fluorescence (TIRF) images were acquired using a Nikon Eclipse Ti fluorescence microscope equipped with a Plan Apo TIRF 100 NA1.49 objective, a Perfect-Focus System, 647 nm diode laser, and an Andor iXon EMCCD DU-897 camera. Before TIRF imaging, differential interference contrast imaging was used to distinguish transfected neurons from glia. Imaging was performed at $37 ^{\circ}$C using objective and stage heaters. Images were background-subtracted and filtered using a Gaussian kernel. Tracking individual fluorophores was performed in MATLAB using the U-track algorithm \cite{jaqaman2008robust}.

\subsection{Quantum dots in the cytoplasm of HeLa cells}
Quantum dots (Qdot 655 ITK Carboxyl, ThermoFisher) were incorporated into
HeLa cells by bead loading, followed by a relaxation time of 1 h before imaging, as previously described \cite{sabri2020elucidating}. Briefly, cells were cultured in Dulbecco’s minimal essential medium with phenol red, supplemented with 10\% fetal calf serum and incubated in culture dishes at $37 ^{\circ}$C in a 5\% CO$_2$ atmosphere. 
Cells were plated on matrigel-coated dishes in Opti-MEM (ThermoFisher). Quantum dots were transferred to the cytoplasm by bead-loading. Imaging was performed 36–48 h after seeding in a custom-built microscope equipped with an Olympus PlanApo 100x NA1.45 objective, a CRISP ASI autofocus system, a back-illuminated Andor iXon DU-888 EMCCD camera, and a 561 nm laser (OBIS, Coherent). The sample was kept at $37 ^{\circ}$C using a Bioptechs stage heater. Images were collected at 10 frames/s. 
Trajectories with at least 100 consecutive positions were obtained with the ImageJ plugin TrackMate \cite{tinevez2017trackmate}. Tracking was performed using the Laplacian-of-Gauss algorithm, and identified particle positions were linked using the simple linear assignment problem tracker.

\subsection{Trajectory segmentation using local convex hull}
Motion changes in particle trajectories were identified using the local convex hull \cite{seg1,sabri2020elucidating}. Trajectories were first rescaled by their average displacement. Then, we employed a sliding window of five trajectory points and calculated within each window the largest diameter $S_d(t)$ of the local convex hull. In order to identify the states with low and high mobility, we selected a threshold ($\eta$) on $S_d$ by using its mean ($\mu$) and standard deviation ($\sigma$). Specifically, we set $\eta=\mu+0.5\sigma$. For any excursion $S_d(t)\geq\eta$, the particle was classified as belonging to the high-mobility state and vice versa. 

\subsection{Turning angles}
We define the turning angles $\theta_\Delta(t)$ at time $t$ and a fixed lag time $\Delta$, from successive displacements
$\delta_\Delta \mathbf{R}(t) \equiv \mathbf{R}(t + \Delta)-\mathbf{R}(t)$ as \cite{sadegh2017plasma}
\begin{align}
   \theta(t)=\cos^{-1}\left(\frac{\delta_\Delta \mathbf{R}(t)\cdot \delta_\Delta \mathbf{R}(t+\Delta)}{\abs{\delta_\Delta \mathbf{R}(t)}\abs{\delta_\Delta \mathbf{R}(t+\Delta)}}\right),
   \label{eq:turning_angle}
\end{align}
where the angles are defined between $0$ and $\pi$. 

\subsection{Numerical simulations}

Whenever full trajectories were required, anisotropic 2D-FBM was simulated using the circulant embedding method that utilizes fast Fourier transform \cite{chan1999simulation}. We simulated 1,000 trajectories of length 4,096. We used the anisotropic 2D-FBM model introduced and described in detail in Supplementary Methods \ref{si:b1}--\ref{si:b3}. We started from two independent FBM components, $R_1(t)$ and $R_2(t)$, with Hurst exponent $H$, but with different diffusivities $D_1$ and $D_2$. We set $D_1 = 1$ and $D_2 = r$, where $r$ is the anisotropy parameter, $0<r\le1$.

To obtain the turning angles look-up tables, we simplified the simulation method. For each value $r$, we used the joint distribution of two consecutive increments of the anisotropic 2D-FBM model. Specifically, for $\Delta = 1$ unit increment step, we considered two consecutive displacement vectors
\begin{align}
\mathbf{U}&=\delta_1\mathbf{\widehat{R}}(t)=\mathbf{\widehat{R}}(t+1)-\mathbf{\widehat{R}}(t),\\ \mathbf{V}&=\delta_1\mathbf{\widehat{R}}(t+1)=\mathbf{\widehat{R}}(t+2)-\mathbf{\widehat{R}}(t+1).
\end{align}
For the non-rotated process, the covariance matrix is 
\begin{align}
\Lambda_r=2\begin{bmatrix} 1 & 0\\ 0 & r \end{bmatrix}.
\end{align}
The covariance between two consecutive one-dimensional FBM increments is $
\gamma_H=2^{2H-1}-1$ and
%
the full covariance matrix of the pair $(\mathbf{U},\mathbf{V})^\mathsf{T}$ is
\begin{align}
\Sigma(r)=\begin{bmatrix} \Lambda_r & \gamma_H\Lambda_r\\ \gamma_H\Lambda_r & \Lambda_r \end{bmatrix}.
\end{align}
For each value of $r$, we sampled $(\mathbf{U},\mathbf{V})^\mathsf{T}$
from the multivariate normal distribution with zero mean and covariance matrix $\Sigma(r)$, where $ \mathbf{U},\mathbf{V}\in\mathbb{R}^2$ represent two consecutive displacement vectors. Then, we computed their corresponding turning angle $
\theta=\cos^{-1}\left(\mathbf{U}\cdot \mathbf{V}/(|\mathbf{U}||\mathbf{V}|)\right).$
For each $r$, the number of simulated angles was matched to the number of empirical turning angles in the corresponding dataset.
Performing simulations in this way is faster than simulating full trajectories of anisotropic 2D-FBM. At the same time, we note that some information is lost, e.g., the structure of dependence in the consecutive turning angles. However, we checked that the obtained distributions remain sufficiently similar for the creation of the look-up tables. 

\backmatter

\bmhead{Data availability}
The datasets generated during and/or analyzed during the current study are available from the corresponding author on reasonable request.

\bmhead{Code availability}
Julia and Python notebooks for numerical simulations, look-up tables generation, and data analysis based on the methods in this article are publicly available on GitHub at \url{https://github.com/MichalBalcerek/Anisotropic-2D-FBM/}.

\bibliography{bibliography}

\bmhead{Acknowledgements}
DK thanks Liz Akin and Michael Tamkun for providing the CD4 and K$_\textrm{v}$1.4 data. AW acknowledges support of the project "Center for Trustworthy Artificial Intelligence for Life Sciences" implemented under the International Research Agendas programme of the Foundation for Polish Science co-financed by the European Union under the European Funds for Smart Economy 2021-2027 (FENG). 

\bmhead{Author contributions}
MB, AW, and DK designed research, MB performed numerical simulations, MB analyzed experimental data and simulations, MB, APP, AW, and DK interpreted results, MB and DK wrote the article with input from all the authors.

\bmhead{Additional information}
\textbf{Competing interests.}
The authors declare no competing interests.

\begin{appendices}
\section{Supplementary information}
\setcounter{figure}{0}
\renewcommand{\thefigure}{S\arabic{figure}}
\renewcommand{\figurename}{Supplementary Fig.}
\renewcommand{\theHfigure}{S\arabic{figure}}
\setcounter{equation}{0}
\renewcommand{\theequation}{S\arabic{equation}}

\begin{figure}
    \centering    \includegraphics[width=1\linewidth]{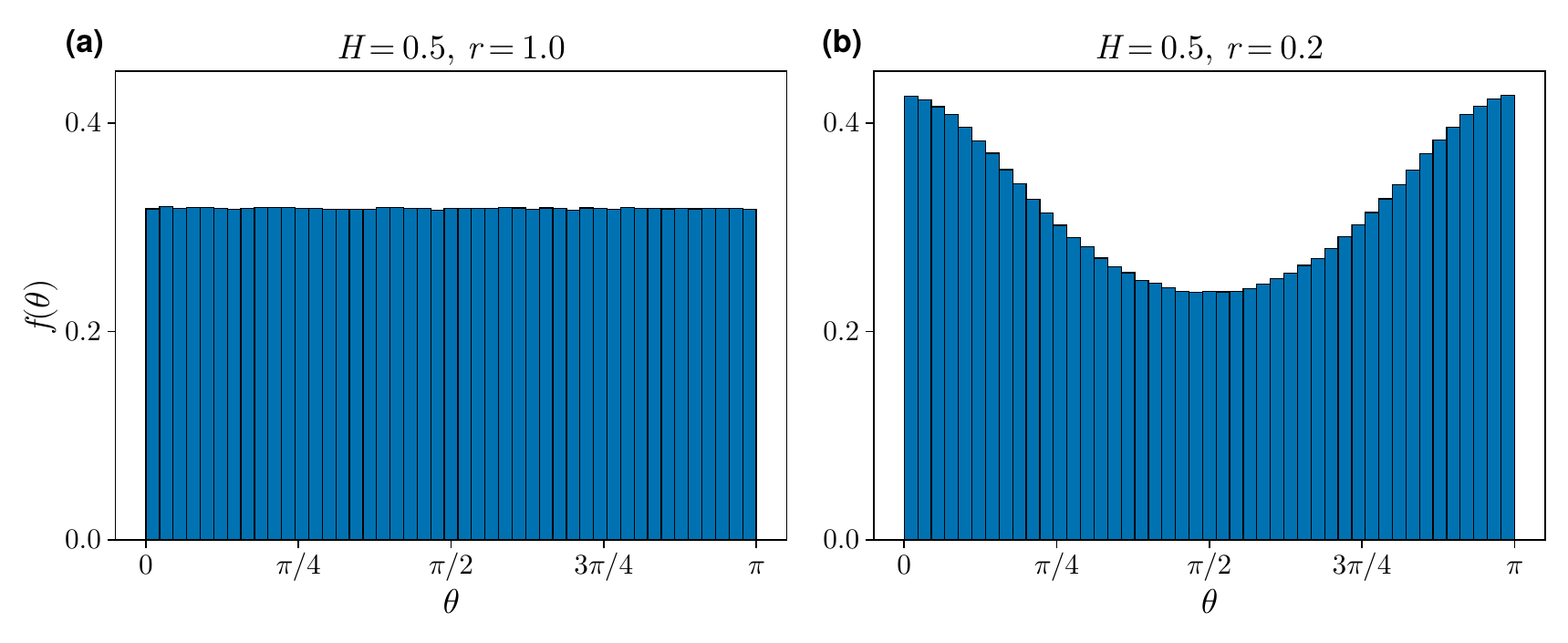}
    \caption{\textbf{Distribution of turning angles in simulated 2D-FBM trajectories with $H=0.5$, i.e., Brownian motion}. (a) Isotropic Brownian motion with $r=1$. (b) Anisotropic Brownian motion with anisotropy ratio $r=0.2$.}
    \label{fig:S1}
\end{figure}

\clearpage

\begin{figure}
    \centering
    \includegraphics[width=1\linewidth]{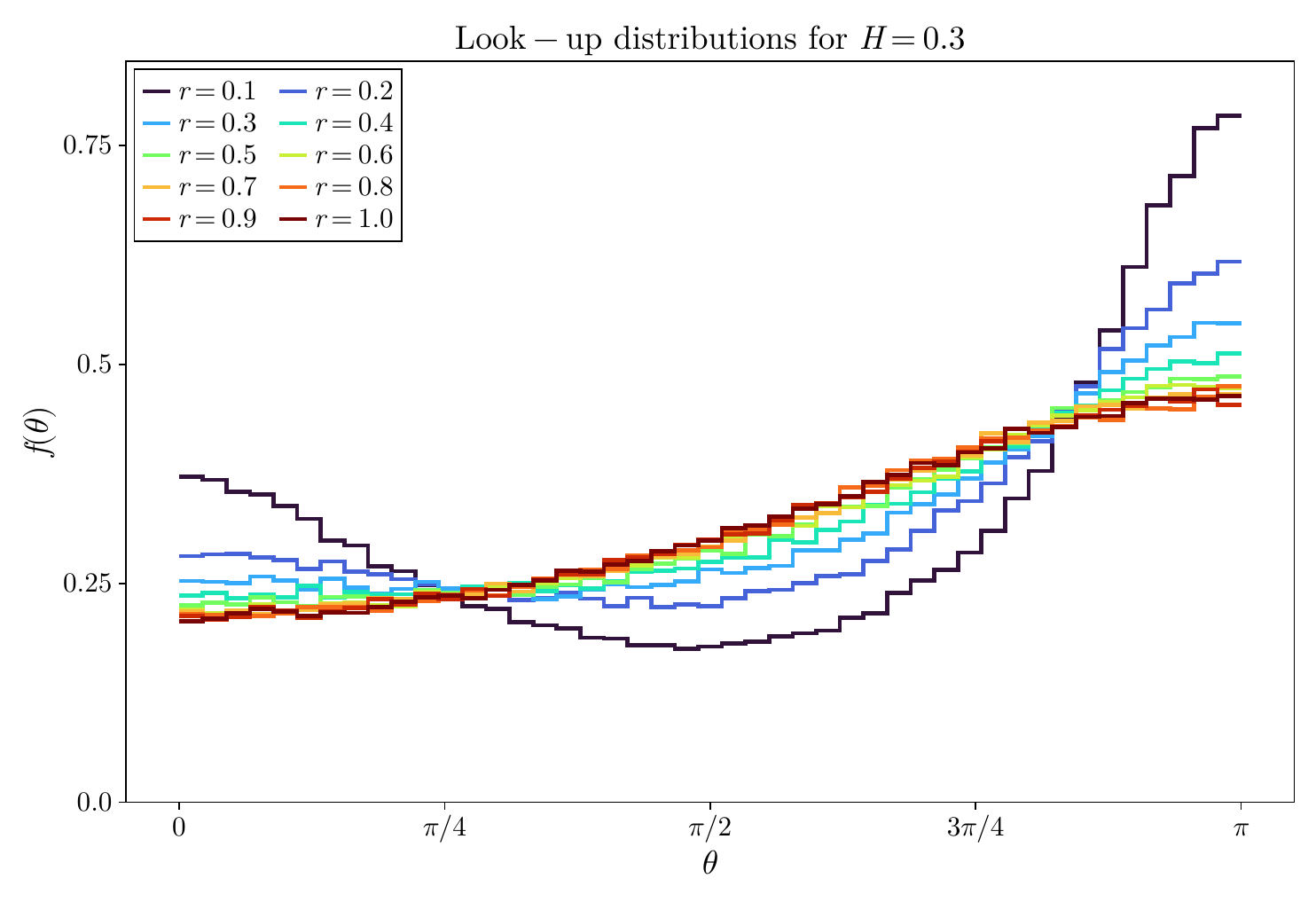}
    \caption{\textbf{Turning-angle distributions used to construct a look-up table.} The distributions were obtained from simulated 2D-FBM trajectories with a fixed Hurst exponent $H=0.3$ and anisotropy ratios $r=0.1,0.2,\ldots,1.0$. The resulting look-up table is used to compare simulated and empirical turning-angle distributions and infer the anisotropy ratio.}
    \label{fig:S2}
\end{figure}
    
\clearpage

\begin{figure}
    \centering
    \includegraphics[width=1\linewidth]{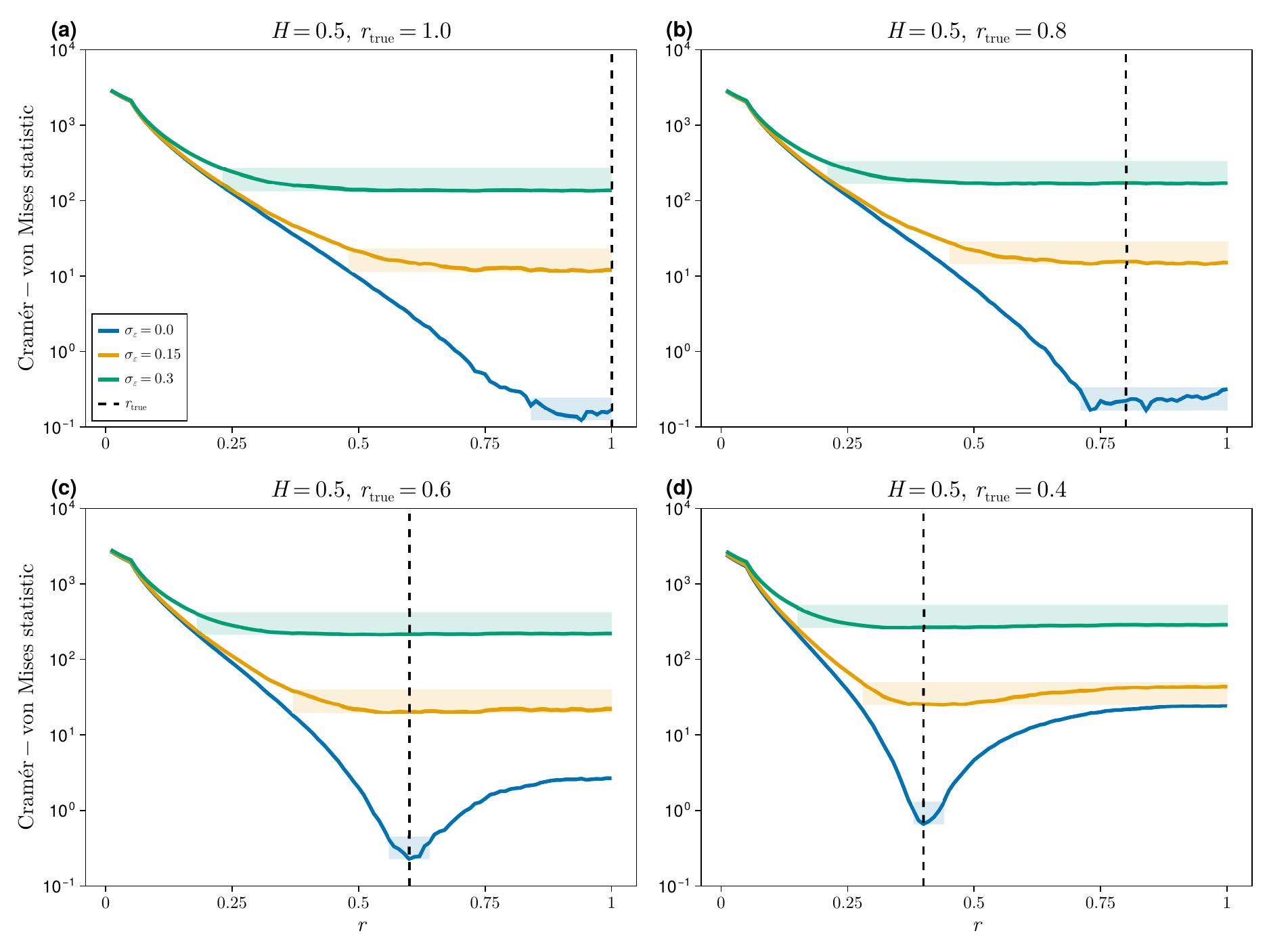}
    \caption{\textbf{Cram\'er--von Mises distance as a function of the anisotropy ratio $r$ for simulated anisotropic 2D {Brownian motion} trajectories.} The realizations have $H=0.5$ and true anisotropy ratios (a) $r_{\mathrm{true}}=1.0$, (b) $r_{\mathrm{true}}=0.8$, (c) $r_{\mathrm{true}}=0.6$, and (d) $r_{\mathrm{true}}=0.4$. Curves correspond to additive Gaussian localization noise with standard deviations $\sigma_\varepsilon=0$ (blue), $0.15$ (yellow), and $0.30$ (green). The vertical dashed lines indicate the true values $r_{\mathrm{true}}$. The shaded regions indicate the uncertainty ranges.}
    \label{fig:S3}
\end{figure}

\clearpage

\begin{figure}
    \centering
    \includegraphics[width=1\linewidth]{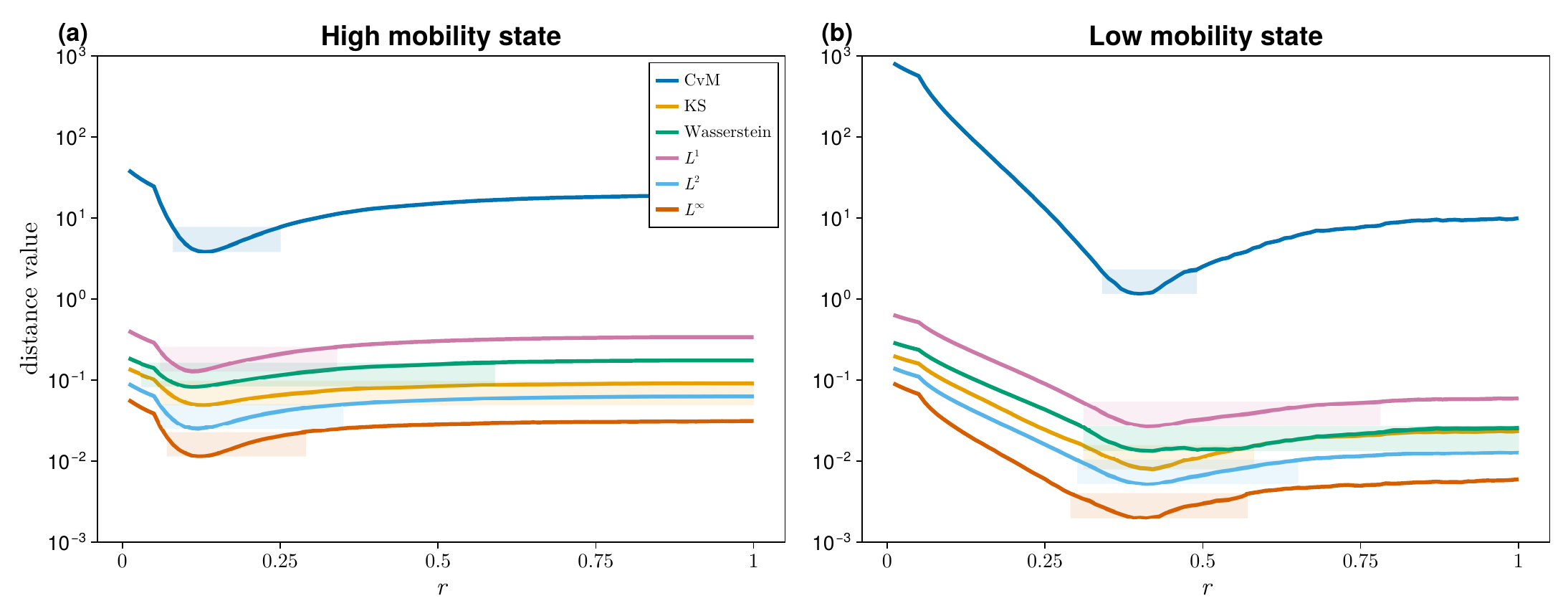}
    \caption{\textbf{All considered distances between the experimental turning-angle distributions and the corresponding simulated look-up tables.} Discrepancy between the experimental and simulated turning-angle distributions as a function of the candidate anisotropy ratio $r$ for the (a) high-mobility and (b) low-mobility states of K$_\textrm{v}$1.4 channels. The distributions are compared using the Cram\'er--von Mises (CvM), Kolmogorov--Smirnov (KS), Wasserstein, $L^1$, $L^2$, and $L^\infty$ distances.}
    \label{fig:S4}
\end{figure}

\clearpage

\begin{figure}
    \centering
    \includegraphics[width=1\linewidth]{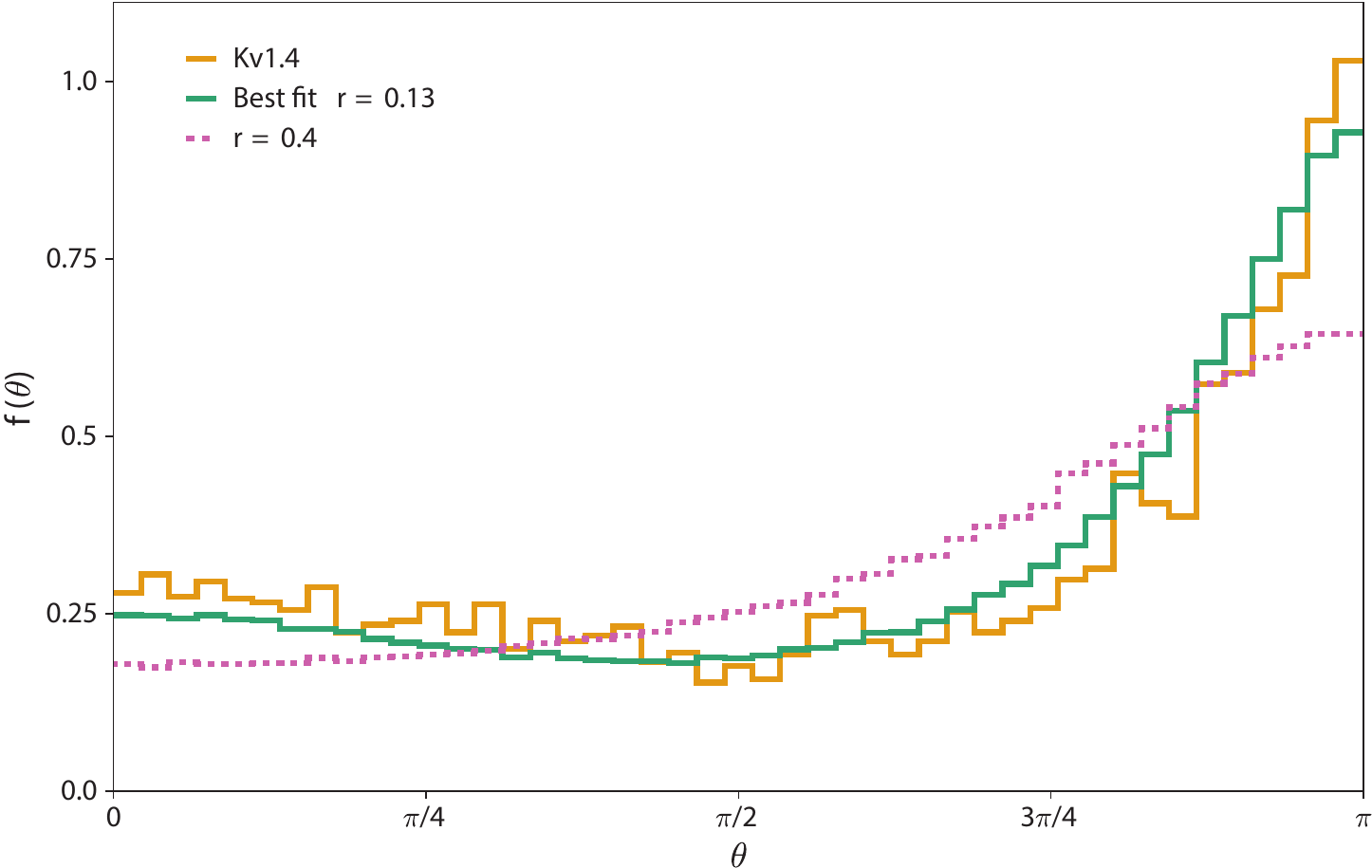}
    \caption{{\textbf{Comparison of turning angle distributions for experimental data and numerical simulations}. The distribution for the high mobility state of K$_\textrm{v}$1.4 is shown together with two distributions from numerical simulations within the uncertainty region of the anisotropy parameter ($r=0.13$) and just outside the interval ($r=0.4$, dashed line).}}
    \label{fig:S5}
\end{figure}

\begin{figure}
    \centering
    \includegraphics[width=1\linewidth]{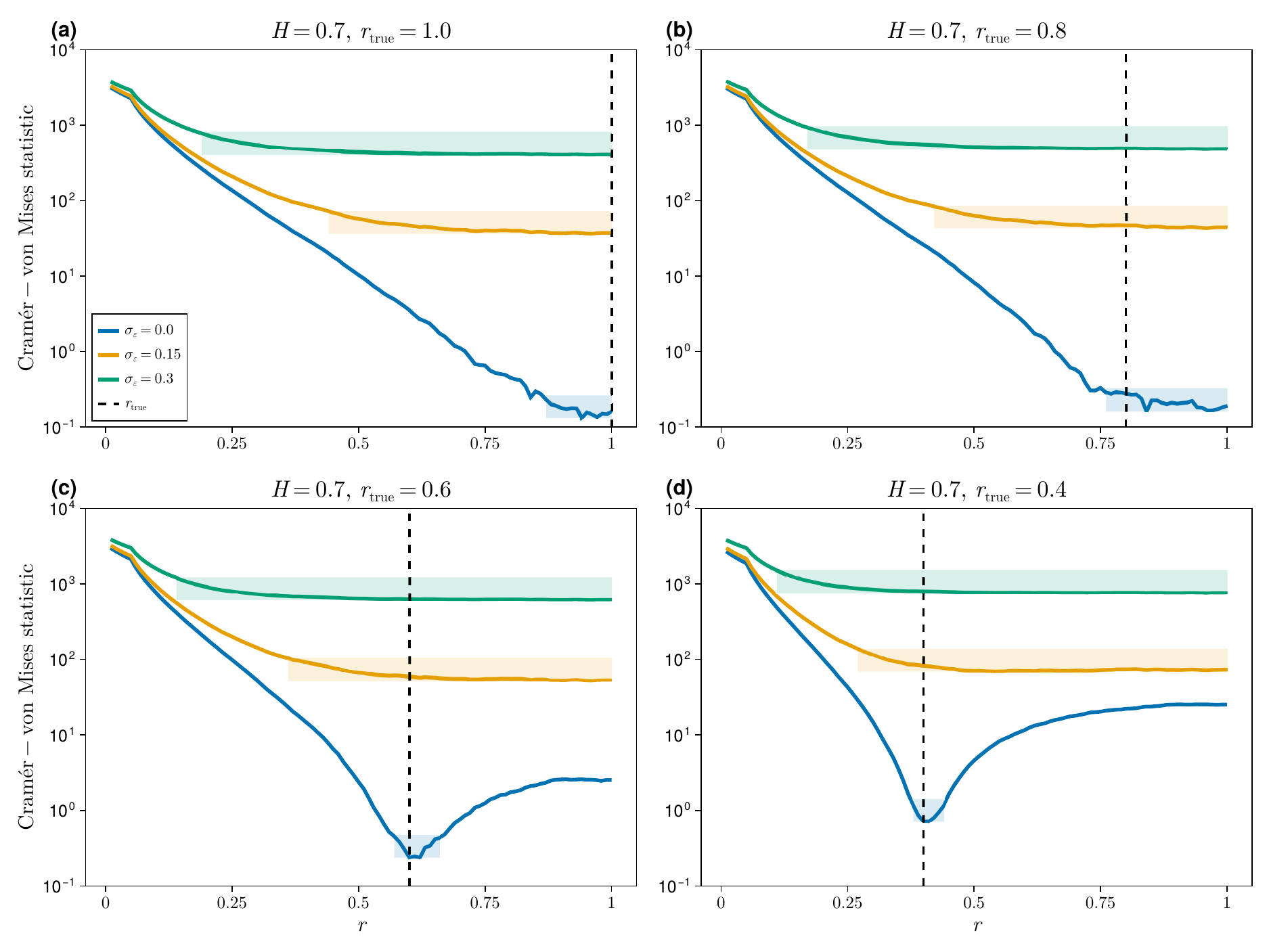}
    \caption{\textbf{Cram\'er--von Mises distance as a function of the anisotropy ratio $r$ for simulated anisotropic 2D-FBM trajectories.} The realizations have $H=0.7$ and true anisotropy ratios (a) $r_{\mathrm{true}}=1.0$, (b) $r_{\mathrm{true}}=0.8$, (c) $r_{\mathrm{true}}=0.6$, and (d) $r_{\mathrm{true}}=0.4$. Curves correspond to additive Gaussian localization noise with standard deviations $\sigma_\varepsilon=0$ (blue), $0.15$ (yellow), and $0.30$ (green). The vertical dashed lines indicate the true values $r_{\mathrm{true}}$, while the shaded regions show the uncertainty ranges.}
    \label{fig:S6}
\end{figure}

\clearpage

\section{Supplementary Methods}

\subsection{Mathematical model for anisotropic two-dimensional Fractional Brownian motion (2D-FBM)}
\label{si:b1}

We consider two independent standard fractional Brownian motions $B_1^H(t)$ and $B_2^H(t)$, $t\geq 0$, with the same Hurst exponent $H_1=H_2=H\in(0,1)$. They may be represented in the causal form 
\begin{align}
B_i^H(t)=\sqrt{a_H}\int_{-\infty}^{t}f\left(s;t,H-\frac12\right)\dd W_i(s),\qquad i=1,2,
\end{align}
where $W_1(t),W_2(t)$ are independent Brownian motions on the real line and
\begin{align}
f\left(s;t,H-\frac12\right)=(t-s)_+^{H-1/2}-(-s)_+^{H-1/2},
\end{align}
with $(x)_+\equiv\max\{x,0\}$.
The normalization constant is \cite{mishura2008stochastic} 
\begin{align}
    a_H=\frac{2H\sin(\pi H) \Gamma(2H)}{\Gamma^2(H+1/2)}
\end{align} 
so that 
\begin{align}
\left\langle B_i^H(t)B_i^H(s)\right\rangle= \frac12\left(|t|^{2H}+|s|^{2H}-|t-s|^{2H}\right), \qquad t, s \geq  0,
\end{align}
for $i=1,2$.

We introduce anisotropy by assigning different generalized diffusion coefficients $D_1$ and $D_2$ to the two independent components,
\begin{align}
\widehat{\mathbf{R}}(t)=\begin{bmatrix}
\widehat{R}_1(t)\\ \widehat{R}_2(t)
\end{bmatrix}=
\begin{bmatrix}
\sqrt{2D_1}B_1^H(t)\\
\sqrt{2D_2}B_2^H(t)
\end{bmatrix},
\label{eq:si_model_def}
\end{align}
where, without loss of generality, $D_1\geq D_2>0$. The anisotropy ratio is defined as
\begin{align}
r={\frac{D_2}{D_1}},\qquad 0<r\leq 1.
\end{align}
Thus, $r=1$ corresponds to isotropic motion, whereas decreasing $r$ indicates increasing anisotropy.
The covariance structure of the process is 
\begin{align}
\boldsymbol{\gamma}(t,s)\equiv\left\langle \widehat{\mathbf{R}}(t)\widehat{\mathbf{R}}(s)^\mathsf{T} \right\rangle= {\boldsymbol{\Lambda}}K_H(t,s),
\end{align}
where $K_H(t, s) =|t|^{2H}+|s|^{2H}-|t-s|^{2H}$ and $\boldsymbol{\Lambda}$ is the diffusion tensor of an anisotropic process with independent components, 
\begin{align}
\boldsymbol{\Lambda}=\begin{bmatrix}D_1&0\\0&D_2\end{bmatrix}=D_1\begin{bmatrix}1&0\\0&r\end{bmatrix}.
\end{align}
Equivalently, 
\begin{align}
\gamma_{11}(t,s)&\equiv\left\langle \widehat{R}_1(t)\widehat{R}_1(s)\right\rangle=D_1\left(|t|^{2H}+|s|^{2H}-|t-s|^{2H}\right),\\
\gamma_{22}(t,s)&\equiv\left\langle \widehat{R}_2(t)\widehat{R}_2(s)\right\rangle=D_2\left(|t|^{2H}+|s|^{2H}-|t-s|^{2H}\right),\\
\gamma_{12}(t,s)&\equiv\left\langle \widehat{R}_1(t)\widehat{R}_2(s)\right\rangle=0.
\end{align}
We consider the same Hurst exponent in both directions so that the process is time-reversible, and their linear combinations remain fractional Brownian motion. Moreover, as a byproduct, we note that its covariance structure remains separable into a common temporal kernel and a spatial diffusion tensor.

We define the increments over a fixed interval $\Delta>0$ as
\begin{align}
\delta_\Delta \widehat{\mathbf{R}}(t) = \widehat{\mathbf{R}}(t+\Delta)-\widehat{\mathbf{R}}(t).
\end{align}
Their covariance function  depends only on the lag time $h$ and is given by
\begin{align}
\mathbf{C}(h)\equiv\left\langle\delta_\Delta\widehat{\mathbf{R}}(t+h)\delta_\Delta\widehat{\mathbf{R}}(t)^\mathsf{T}\right\rangle=G_{H,\Delta}(h)\boldsymbol{\Lambda},
\end{align}
where
\begin{align}
G_{H,\Delta}(h)=|h+\Delta|^{2H}+|h-\Delta|^{2H}-2|h|^{2H}, \qquad h \in \R.
\end{align}
Consequently its elements are
\begin{align}
C_{11}(h)&\equiv\left\langle\delta_\Delta \widehat{R}_1(t+h)\delta_\Delta \widehat{R}_1(t)\right\rangle=D_1G_{H,\Delta}(h),\\
C_{22}(h)&\equiv\left\langle\delta_\Delta \widehat{R}_2(t+h)\delta_\Delta \widehat{R}_2(t)\right\rangle=D_2G_{H,\Delta}(h),\\
C_{12}(h)&\equiv\left\langle\delta_\Delta \widehat{R}_1(t+h)\delta_\Delta \widehat{R}_2(t)\right\rangle=0.
\end{align}
To calculate the second moment of the increments we take $h=0$, and therefore we have 
\begin{align}
\left\langle \left(\delta_\Delta \widehat{R}_1(t)\right)^2\right\rangle &=2D_1\Delta^{2H},\\ 
\left\langle \left(\delta_\Delta \widehat{R}_2(t)\right)^2\right\rangle &=2D_2\Delta^{2H}.
\end{align}

Let us now consider the anisotropic 2D-FBM process after a deterministic rotation by an angle $\varphi$,
\begin{align}
\mathbf{R}(t)=\mathbf{A}_\varphi\mathbf{\widehat{R}}(t),
\end{align}
where $\mathbf{A}_\varphi$ is a rotation matrix,
\begin{align}
\mathbf{A}_\varphi=\begin{bmatrix}\cos\varphi&-\sin\varphi\\\sin\varphi& \phantom{-} \cos\varphi\end{bmatrix}.
\end{align}
The diffusion tensor of the rotated process is 
\begin{align}
\mathbf{D}=\mathbf{A}_\varphi\boldsymbol{\Lambda}\mathbf{A}_\varphi^\mathsf{T}=
\begin{bmatrix}D_1\cos^2\varphi+D_2\sin^2\varphi&(D_1-D_2)\sin\varphi\cos\varphi\\(D_1-D_2)\sin\varphi\cos\varphi&D_1\sin^2\varphi+D_2\cos^2\varphi\end{bmatrix}.
\end{align}
Equivalently, using $D_2=rD_1$ we have
\begin{align}
\mathbf{D}=D_1
\begin{bmatrix}
\cos^2\varphi+r\sin^2\varphi&(1-r)\sin\varphi\cos\varphi\\
(1-r)\sin\varphi\cos\varphi&\sin^2\varphi+r\cos^2\varphi
\end{bmatrix}.
\end{align}
The covariance structures of the rotated process and its increments are therefore
\begin{align}
\boldsymbol{\gamma}(t,s)&= K_H(t,s)\mathbf{D},\\ 
\mathbf{C}(h) &= G_{H,\Delta}(h)\mathbf{D}.
\end{align}
Componentwise, the covariance functions of the rotated process are
\begin{align}
\left\langle X(t)X(s)\right\rangle&=(D_1\cos^2\varphi+D_2\sin^2\varphi)K_H(t,s),\\
\left\langle Y(t)Y(s)\right\rangle&=(D_1\sin^2\varphi+D_2\cos^2\varphi)K_H(t,s),\\
\left\langle X(t)Y(s)\right\rangle&=(D_1-D_2)\sin\varphi\cos\varphi\,K_H(t,s).
\end{align}
Similarly, the covariance functions of the increments are
\begin{align}
C_{xx}(h)&=\left\langle\delta_\Delta X(t+h)\delta_\Delta X(t)\right\rangle=(D_1\cos^2\varphi+D_2\sin^2\varphi)G_{H,\Delta}(h),\\
C_{yy}(h)&=\left\langle\delta_\Delta Y(t+h)\delta_\Delta Y(t)\right\rangle=(D_1\sin^2\varphi+D_2\cos^2\varphi)G_{H,\Delta}(h),\\
C_{xy}(h)&=\left\langle\delta_\Delta X(t+h)\delta_\Delta Y(t)\right\rangle=(D_1-D_2)\sin\varphi\cos\varphi\,G_{H,\Delta}(h).
\end{align}
Thus, rotation of an anisotropic process produces a non-zero cross-covariance between the components, even though the intrinsic components are independent. The cross-covariance vanishes when $D_1=D_2$, when $\varphi$ is an integer multiple of $\pi/2$, or when the observation axes coincide with the intrinsic principal axes.

\subsection{Relations between the  parametrizations of 2D-FBM}
\label{si:b2}
\subsubsection{Transformation from correlated-component \texorpdfstring{$(D_x,D_y,\rho)$}{(Dx, Dy, rho)} to principal-axis \texorpdfstring{$(D_1,D_2,\varphi)$}{(D1, D2, phi)}
 parametrizations of 2D-FBM} 
\label{si:b21}
In our previous work\cite{balcerek2026two}, 2D-FBM was described directly in an arbitrary frame of reference using two correlated components. In that notation, the diffusion tensor has the form
\begin{align}
\mathbf{D}=\begin{bmatrix}D_x&\rho\sqrt{D_xD_y}\\\rho\sqrt{D_xD_y}&D_y\end{bmatrix},
\end{align}
where $D_x,D_y>0$ are general diffusion coefficients
and $\rho\in[-1,1]$ is the corresponding correlation coefficient. Equivalently, the underlying Wiener processes of the two components are correlated and satisfy
\begin{align}
\left\langle \dd \widetilde W_x(t)\dd \widetilde W_y(t)\right\rangle=\rho\dd t.
\end{align}
The covariance matrices of the process and its increments are then 
\begin{align}
{\boldsymbol{\gamma}}(t,s)=K_H(t,s)\mathbf{D},\qquad {\mathbf{C}}(h)=G_{H,\Delta}(h)\mathbf{D}.
\end{align}
Componentwise, this gives 
\begin{align}
\left\langle X(t)X(s)\right\rangle&=D_xK_H(t,s),\\
\left\langle Y(t)Y(s)\right\rangle&=D_yK_H(t,s),\\
\left\langle X(t)Y(s)\right\rangle&=\rho\sqrt{D_xD_y}\,K_H(t,s).
\end{align}

This approach describes the same class of models as the one presented in the main text. Namely, since $\mathbf{D}$ is symmetric and positive semidefinite, it can be diagonalized by a planar rotation,
\begin{align}
\mathbf{D}=A_{\varphi_0}\begin{bmatrix}\lambda_+&0\\0&\lambda_-\end{bmatrix}A_{\varphi_0}^\mathsf{T},
\end{align}
where
\begin{align}
\lambda_\pm=\frac12\left(D_x+D_y\pm\sqrt{(D_x-D_y)^2+4\rho^2D_xD_y}\right)
\end{align}
are its eigenvalues. The orientation of the principal axes is chosen as
\begin{align}
\varphi_0=\frac{1}{2}\operatorname{atan2}\left(2\rho\sqrt{D_xD_y},D_x-D_y\right),
\end{align}
where $\operatorname{atan2}$ is the 2-argument arctangent function. In the special case where $\rho=0, D_x=D_y$, i.e., an isotropic process with independent components, any rotation will result in the same model. For this reason, we take $\operatorname{atan2}(0,0)=0$. Identifying
\begin{align}
D_1=\lambda_+,\qquad D_2=\lambda_-,
\end{align}
gives the intrinsic independent-component representation introduced in Eq. \eqref{eq:si_model_def}, with
\begin{align}
r=\frac{D_2}{D_1}=\frac{\lambda_-}{\lambda_+}.
\end{align}
Thus, $r$ measures the ratio of the diffusivities along the principal axes, whereas $\rho$ measures the correlation between components in a specific trajectory. Finally, we note that the parameter $r$ is invariant under rotations, while $\rho$ is not.
In the special case in which the observed marginal diffusivities are equal, $D_x=D_y=D$, the eigenvalues reduce to
\begin{align}
\lambda_\pm=D(1\pm|\rho|).
\end{align}
Consequently, when $D_x=D_y$,
\begin{align}
r=\frac{1-|\rho|}{1+|\rho|}.
\end{align}
 
\subsubsection{Inverse transform: from principal-axis \texorpdfstring{$(D_1,D_2,\varphi)$}{(D1, D2, phi)} to correlated-component \texorpdfstring{$(D_x,D_y,\rho)$}{(Dx, Dy, rho)} parametrizations}
\label{si:b22}

The diffusion coefficients in the rotated coordinate system are 
\begin{align}
D_x=D_1\cos^2\varphi+D_2\sin^2\varphi,\qquad D_y=D_1\sin^2\varphi+D_2\cos^2\varphi,
\end{align}
and the correlation coefficient is
\begin{align}
\rho=\frac{(D_1-D_2)\sin\varphi\cos\varphi}{\sqrt{(D_1\cos^2\varphi+D_2\sin^2\varphi)(D_1\sin^2\varphi+D_2\cos^2\varphi)}}.
\end{align}
Using $D_2=rD_1$, this expression becomes
\begin{align}
\rho=\frac{(1-r)\sin\varphi\cos\varphi}{\sqrt{(\cos^2\varphi+r\sin^2\varphi)(\sin^2\varphi+r\cos^2\varphi)}}.
\end{align}
Hence, a non-zero value of $\rho$ can arise solely from observing independent but anisotropic components along an orientation different from their principal directions.

\subsection{Anisotropic model under random rotations}
\label{si:b3}
In many physical scenarios, the orientation of the intrinsic principal axes is not fixed but varies from trajectory to trajectory. We describe this situation by treating the rotation angle as a random variable $\Phi$ distributed uniformly on the interval $(0,2\pi)$, independent of the underlying fractional Brownian motions. Conditional on $\Phi=\varphi$, each realization remains a 2D-FBM with $\mathbf{D}$. When the orientation changes from trajectory to trajectory, the ensemble obtained by averaging over $\Phi$ is generally a mixture of Gaussian processes for $r<1$.
For a uniformly distributed $\Phi$, we have 
\begin{align}
\langle \mathbf{D} \rangle_\Phi =\frac{D_1+D_2}{2}\mathbf{I}_2,
\end{align}
and, in particular,
\begin{align}
\langle C_{xy}(h)\rangle_\Phi=0
\end{align}
for every lag time $h$. Therefore, covariance and cross-covariance functions calculated after pooling randomly oriented trajectories appear isotropic, even though each individual realization may have anisotropy $r<1$.

An important geometric observable in this context is the sequence of turning angles. Let $U,V\in\mathbb{R}^2$ denote two consecutive displacement vectors. Their turning angle is
\begin{align}
\theta(U,V)=\cos^{-1}\left(\frac{U\cdot V}{|U||V|}\right).
\end{align}
Since rotations preserve scalar products and Euclidean norms,
\begin{align}
\theta(A_\varphi U,A_\varphi V)=\theta(U,V).
\end{align}
Consequently, the distribution of turning angles depends on the Hurst exponent $H$ and the anisotropy ratio $r$, but not on the deterministic or random orientation of the trajectory. 

\subsection{Estimation protocol based on the distribution of turning angles}
\label{si:b4}
\begin{figure}
    \centering
    \includegraphics[width=1\linewidth]{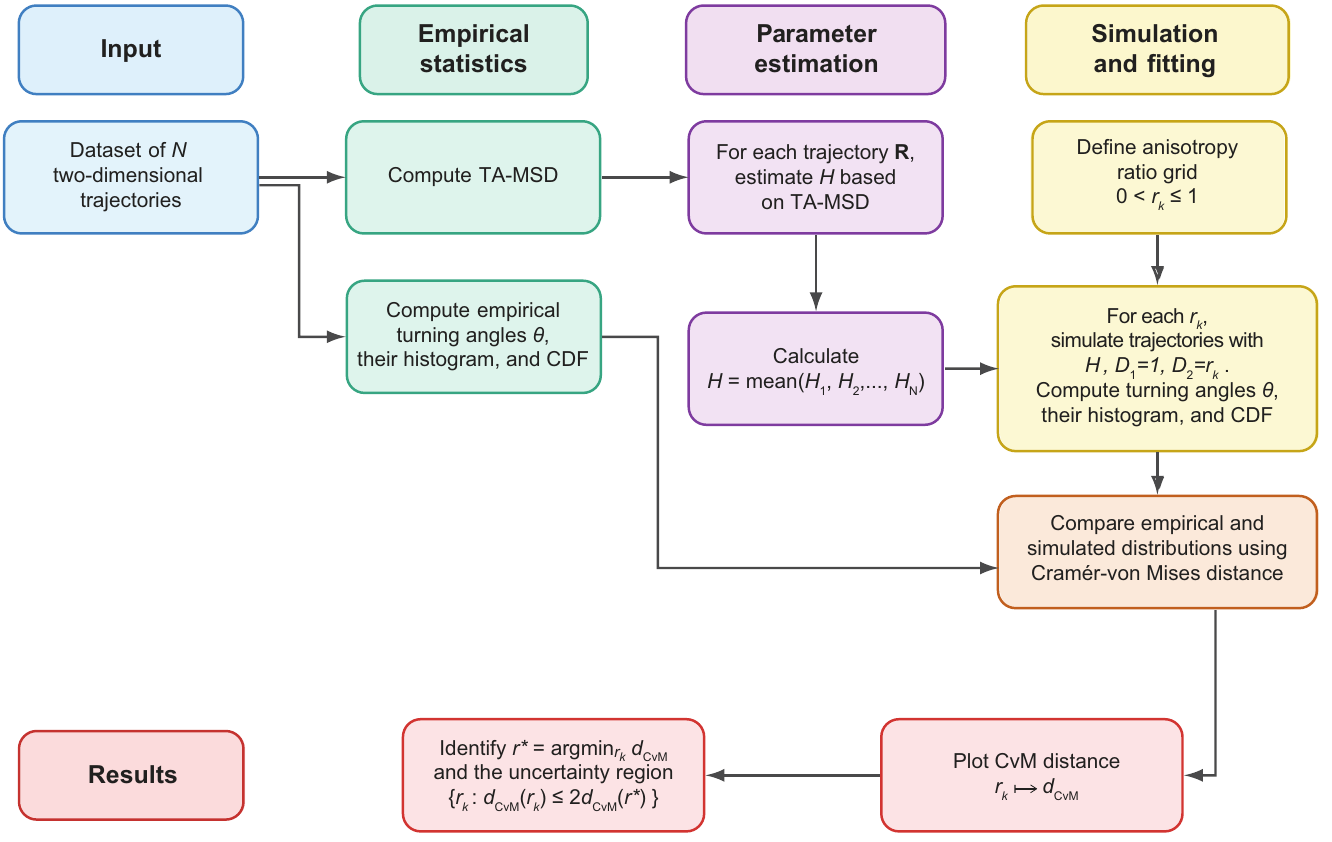}
    \caption{\textbf{Workflow for estimating the anisotropy ratio $r$ from trajectory data.} Starting from a dataset, the empirical distribution of turning angles is computed. The Hurst exponent (H) is estimated for each trajectory, and the mean value is used. A grid of candidate anisotropy ratios $r_k$ is defined in the interval $(0,1]$. Anisotropic FBM trajectories are simulated for each $r_k$. The turning-angle distributions are calculated and compared with the empirical distribution using the Cramér-von Mises (CvM) distance. The anisotropy ratio $r$ is estimated as the value that minimizes CvM distance.}
    \label{fig:algorithm}
\end{figure}

Supplementary Fig.~\ref{fig:algorithm} summarizes the procedure used to estimate the anisotropy ratio $r$. For each experimental dataset or mobility state, we first calculate the empirical turning angles and estimate the Hurst exponent $H$ from the two-dimensional mean squared displacement. We then generate look-up tables of turning-angle distributions for the estimated $H$ and a grid of candidate anisotropy ratios $r_k\in(0,1]$.

We measure the discrepancy between the
empirical and simulated turning-angle distributions using a modification of the two-sample Cram\'er--von Mises criterion \cite{anderson1962distribution}. 
Let $\theta_1^{\mathrm{emp}},\ldots,\theta_n^{\mathrm{emp}}$ denote the empirical turning angles and let $\theta_1^{(r_k)},\ldots,\theta_m^{(r_k)}$ denote the simulated turning angles corresponding to the candidate anisotropy ratio $r_k$. Their empirical cumulative distribution functions are 
\begin{align}
F_n(\theta) &=\frac{1}{n} \sum_{i=1}^{n}\mathbf{1}_{\{\theta_i^{\mathrm{emp}}\leq\theta\}},\\
G_m^{(r_k)}(\theta)&= \frac{1}{m}\sum_{j=1}^{m} \mathbf{1}_{\{\theta_j^{(r_k)}\leq\theta\}}.
\end{align}
Let
\begin{align}
H_{n,m}^{(r_k)}(\theta) = \frac{nF_n(\theta)+mG_m^{(r_k)}(\theta)}{n+m}
\end{align}
be the empirical cumulative distribution function of the pooled sample.
The two-sample Cram\'er--von Mises criterion is
\begin{align}
d_{\mathrm{CvM}}(r_k) = \frac{nm}{n+m} \int_{0}^{\pi}
\left[ F_n(\theta)-G_m^{(r_k)}(\theta) \right]^2
\dd H_{n,m}^{(r_k)}(\theta).
\end{align}
Equivalently, if $z_1\leq\cdots\leq z_{n+m}$ denotes the ordered pooled
sample, including repeated observations, then
\begin{align}
d_{\mathrm{CvM}}(r_k) = \frac{nm}{(n+m)^2}
\sum_{\ell=1}^{n+m} \left[F_n(z_\ell)-G_m^{(r_k)}(z_\ell)
\right]^2.
\end{align}
The resulting distance curve can be additionally smoothed over neighboring values (via averaging of 9 neighbours) of $r_k$, and the estimated anisotropy ratio is defined as
\begin{align}
r^*=\operatorname{argmin}_{r_k}\widetilde{d}_{\mathrm{CvM}}(r_k),
\end{align}
where $\widetilde{d}_{\mathrm{CvM}}$ denotes the smoothed distance. The uncertainty range is finally defined as
\begin{align}
\left\{r_k:\widetilde{d}_{\mathrm{CvM}}(r_k)\leq 2\widetilde{d}_{\mathrm{CvM}}(r^*)\right\}.
\end{align}

\subsection{Alternative distance measures to identify \texorpdfstring{$r$}{r}}
\label{si:b5}
In addition to the Cram\'er--von Mises distance introduced in Supplementary Methods \ref{si:b4}, we assessed the robustness of the anisotropy ratio $r$ estimation using several alternative distance measures. We retain the notation from the previous section: $F_n$ denotes the empirical cumulative distribution function of the measured turning angles, whereas $G_m^{(r_k)}$ denotes the empirical cumulative distribution function of the simulated turning angles for the candidate anisotropy ratio $r_k$.

For histogram-based comparisons, we partition $[0,\pi]$ into $B$ equal-width bins $I_1,\ldots,I_B$. In the numerical implementation, we used $B=45$. Let
\begin{align}
p_i&=\frac{1}{n}\sum_{j=1}^{n}\mathbf{1}_{\{\theta_j^{\mathrm{emp}}\in I_i\}},&
q_i^{(r_k)}&=\frac{1}{m}\sum_{j=1}^{m}\mathbf{1}_{\{\theta_j^{(r_k)}\in I_i\}}
\end{align}
denote the empirical and simulated probability masses in the $i$-th bin. We consider the distances
\begin{align}
d_{L^1}(r_k)&=\sum_{i=1}^{B}\left|p_i-q_i^{(r_k)}\right|,\\
d_{L^2}(r_k)&=\left[\sum_{i=1}^{B}\left(p_i-q_i^{(r_k)}\right)^2\right]^{1/2},\\
d_{L^\infty}(r_k)&=\max_{1\leq i\leq B}\left|p_i-q_i^{(r_k)}\right|.
\end{align}
We also consider two distances based directly on the empirical cumulative distribution functions and, therefore, independent of the histogram binning. The Kolmogorov--Smirnov distance is defined as \cite{kolmogorov49}
\begin{align}
d_{\mathrm{KS}}(r_k) = \sup_{\theta\in[0,\pi]} \left|F_n(\theta)-G_m^{(r_k)}(\theta)\right|.
\end{align}
Finally, the Wasserstein distance of order one is given by\cite{vallender1974calculation}
\begin{align}
d_{\mathrm{W}}(r_k) = \int_{0}^{\pi} \left|F_n(\theta)-G_m^{(r_k)}(\theta)\right|\dd\theta.
\end{align}

For each distance $d$, a corresponding anisotropy estimate is defined as 
\begin{align}
r_d^*=\operatorname{argmin}_{r_k}d(r_k).
\end{align}
Regardless of the choice of the distance, we observed that their minima occur in similar ranges of $r$ for both the simulated and the experimental data, indicating that the inferred anisotropy is not specific to the use of the Cram\'er--von Mises distance; see Supplementary Fig. \ref{fig:S4} for the case of K$_\textrm{v}$1.4 data.

\end{appendices}

\end{document}